\documentclass[twocolumn,preprintnumbers,superscriptaddress,amsmath,amssymb,prb]{revtex4-1}
\usepackage{braket}
\usepackage{miller}
\usepackage{changes}

\usepackage[textsize=tiny]{todonotes}
\usepackage{graphicx}

\usepackage[utf8]{inputenc}
\usepackage{multirow}
\usepackage{float}
\usepackage{bm}
\usepackage{verbatim}
\usepackage{xfrac}
\usepackage{array}
\usepackage{siunitx}
\usepackage{ulem}

\begin{document}
	
	\title{Optically controlling the emission chirality of microlasers}

	\author{N. Carlon Zambon}
	\thanks{These authors contributed equally to this work.}
	\affiliation{Centre de Nanosciences et de Nanotechnologies (C2N), CNRS - Université Paris-Sud / Paris-Saclay, Marcoussis, France}

	\author{P. St-Jean}
	\thanks{These authors contributed equally to this work.}
	\affiliation{Centre de Nanosciences et de Nanotechnologies (C2N), CNRS - Université Paris-Sud / Paris-Saclay, Marcoussis, France}
	
	\author{M. Mili\'cevi\'c}
	\affiliation{Centre de Nanosciences et de Nanotechnologies (C2N), CNRS - Université Paris-Sud / Paris-Saclay, Marcoussis, France}
	
	\author{A. Lemaître}
	\affiliation{Centre de Nanosciences et de Nanotechnologies (C2N), CNRS - Université Paris-Sud / Paris-Saclay, Marcoussis, France}
	
	\author{A. Harouri}
	\affiliation{Centre de Nanosciences et de Nanotechnologies (C2N), CNRS - Université Paris-Sud / Paris-Saclay, Marcoussis, France}
	
	\author{L. Le Gratiet}
	\affiliation{Centre de Nanosciences et de Nanotechnologies (C2N), CNRS - Université Paris-Sud / Paris-Saclay, Marcoussis, France}
	
	\author{O. Bleu}
	\affiliation{Institut Pascal, PHOTON-N2, CNRS - University Clermont Auvergne, Clermont-Ferrand, France.}
	
	\author{D. D. Solnyshkov}
	\affiliation{Institut Pascal, PHOTON-N2, CNRS - University Clermont Auvergne, Clermont-Ferrand, France.}
	
	\author{G. Malpuech}
	\affiliation{Institut Pascal, PHOTON-N2, CNRS - University Clermont Auvergne, Clermont-Ferrand, France.}
	
	\author{I. Sagnes}
	\affiliation{Centre de Nanosciences et de Nanotechnologies (C2N), CNRS - Université Paris-Sud / Paris-Saclay, Marcoussis, France}
	
	\author{S. Ravets}
	\affiliation{Centre de Nanosciences et de Nanotechnologies (C2N), CNRS - Université Paris-Sud / Paris-Saclay, Marcoussis, France}
	
	\author{A. Amo}
	\affiliation{Physique des Lasers, Atomes et Molécules (PhLAM), CNRS - Université de Lille, Lille, France}
	
	\author{J. Bloch}
	\affiliation{Centre de Nanosciences et de Nanotechnologies (C2N), CNRS - Université Paris-Sud / Paris-Saclay, Marcoussis, France}

	\begin{abstract}
	Orbital angular momentum (OAM) carried by helical light beams is an unbounded degree of freedom of photons that offers a promising playground in modern photonics. So far, integrated sources of coherent light carrying OAM are based on resonators whose design imposes a single, non-tailorable chirality of the wavefront (i.e. clockwise or counter-clockwise vortices). Here, we propose and demonstrate the realization of an integrated microlaser where the chirality of the wavefront can be optically controlled. Importantly, the scheme that we use, based on an effective spin-orbit coupling of photons in a semiconductor microcavity, can be extended to different laser architectures, thus paving the way to the realization of a new generation of OAM microlasers with tunable chirality.
	
	\end{abstract}
	
	\maketitle
	
	\setcounter{topnumber}{2}
	\setcounter{bottomnumber}{2}
	\setcounter{totalnumber}{4}
	\renewcommand{\topfraction}{0.85}
	\renewcommand{\bottomfraction}{0.85}
	\renewcommand{\textfraction}{0.15}
	\renewcommand{\floatpagefraction}{0.7}	
	

	
	Harnessing the physical properties of light, e.g. its frequency, amplitude, wavevector and angular momentum, is ubiquitous in photonic technologies. Among these various degrees of freedom, angular momentum emerging from the spin moment of photons (related to their circular polarization) has been proven to be extremely powerful as it can be easily controlled with linear optical elements such as wave plates and polarizers.
	
	Since the pioneering work of Allen et al. \cite{Allen1992}, it is now well-known that light's angular momentum is not restricted to the spin moment of photons. It can also emerge, in the paraxial regime, by structuring helical phase fronts $e^{i\ell\phi}$, where the quantum number $\ell$ describes the number of times the phase of the wavefront winds around the direction of propagation within an optical period. The most notable asset of this degree of freedom, usually referred to as the orbital angular momentum (OAM), is that, contrary to its spin counterpart which is restricted to values of $\pm\hbar$, it is theoretically unbounded; e.g. generation of light vortices carrying more than $\mathrm{10^{4}}$ quanta of OAM has recently been demonstrated \cite{Fickler2016}.
	
	Over the last decade, numerous proposals and demonstrations that take profit of this unbounded Hilbert space have emerged. For instance, it was acknowledged that these higher-dimensional quantum states could offer a drastically enhanced information density, both in classical \cite{Gibson2004, Wang2012, Bozinovic2013, Willner2015} and quantum\cite{Groblacher2006, Vallone2014, Sit2017, Wang2018, Erhard2018} communication channels; as well, they could allow improving the resilience against noise and eavesdropping of quantum communication protocols\cite{Kaszlikowski2000, Cerf2002}. Moreover, the ability to transfer large quanta of OAM to massive objects has lead to the development of novel techniques in optical manipulation\cite{Grier2003, Padgett2011, Gao2017} and in optomechanics\cite{Aspelmeyer2014}. From a fundamental point of view, generating and entangling quantum states with such arbitrarily large quantum numbers has been demonstrated to be a very promising avenue for investigating the foundations of quantum mechanics\cite{Collins2002, Fickler2012, Fickler2016}.
	
	The growing interest in this degree of freedom of light calls for the development of coherent sources carrying well-defined and tunable OAM. One possible strategy that has been extensively explored is to shape the phase front of paraxial beams with bulk devices such as spiral phase plates\cite{Simpson1997, Schemmel2014}, spatial light modulators\cite{Wang2012, Bozinovic2013, Mirhosseini2013}, nanostructured metasurfaces\cite{Lin2014, Arbabi2015, Devlin2017, Xu2018} or q-plates\cite{Marrucci2006,Karimi2009}; other strategies based on polariton condensation in chiral optical traps have also been demonstrated\cite{Dall2014, Gao2018}. Although these approaches have the advantage of being extremely versatile, allowing to generate high order vortices, they remain extremely difficult to integrate \textit{on-a-chip}.
	
	Therefore, recent demonstrations of integrated OAM lasers based on ring resonators\cite{Miao2016, Peng2016} are very promising. However, it is very challenging in these integrated devices to break the mirror symmetry between clockwise (CW) and counter-clockwise (CCW) propagating modes, which is necessary to generate an emission carrying a net OAM. So far, this difficulty has been succesfully overcome by engineering chiral resonators, e.g. by tuning the gain and loss around the resonator, but the scalability of this approach is strongly limited because the engineering of the devices imposes a given, non-tailorable chirality to the lasing mode: each device can generate only either a CW or CCW vortex.
	
	In this work, we propose and demonstrate a novel scheme to achieve OAM lasing in a fully integrated device where the chirality of the emission (i.e. CW or CCW vortices) can be optically controlled. Rather than relying on the engineering of a chiral resonator, we take profit of the spin-orbit coupling of photons\cite{Sala2015, Cardano2015} confined in a ring resonator with discrete rotational symmetry. This allows to optically break time-reversal symmetry by spin-polarizing the gain medium. Here, we show that this can be achieved with a circularly polarized optical pump, and that the chirality of the emission can be controlled solely by tuning the pump polarization.\\
	
	\begin{figure*}
		\centering
		\includegraphics[trim=0cm 0cm 0cm 0cm, width=150mm]{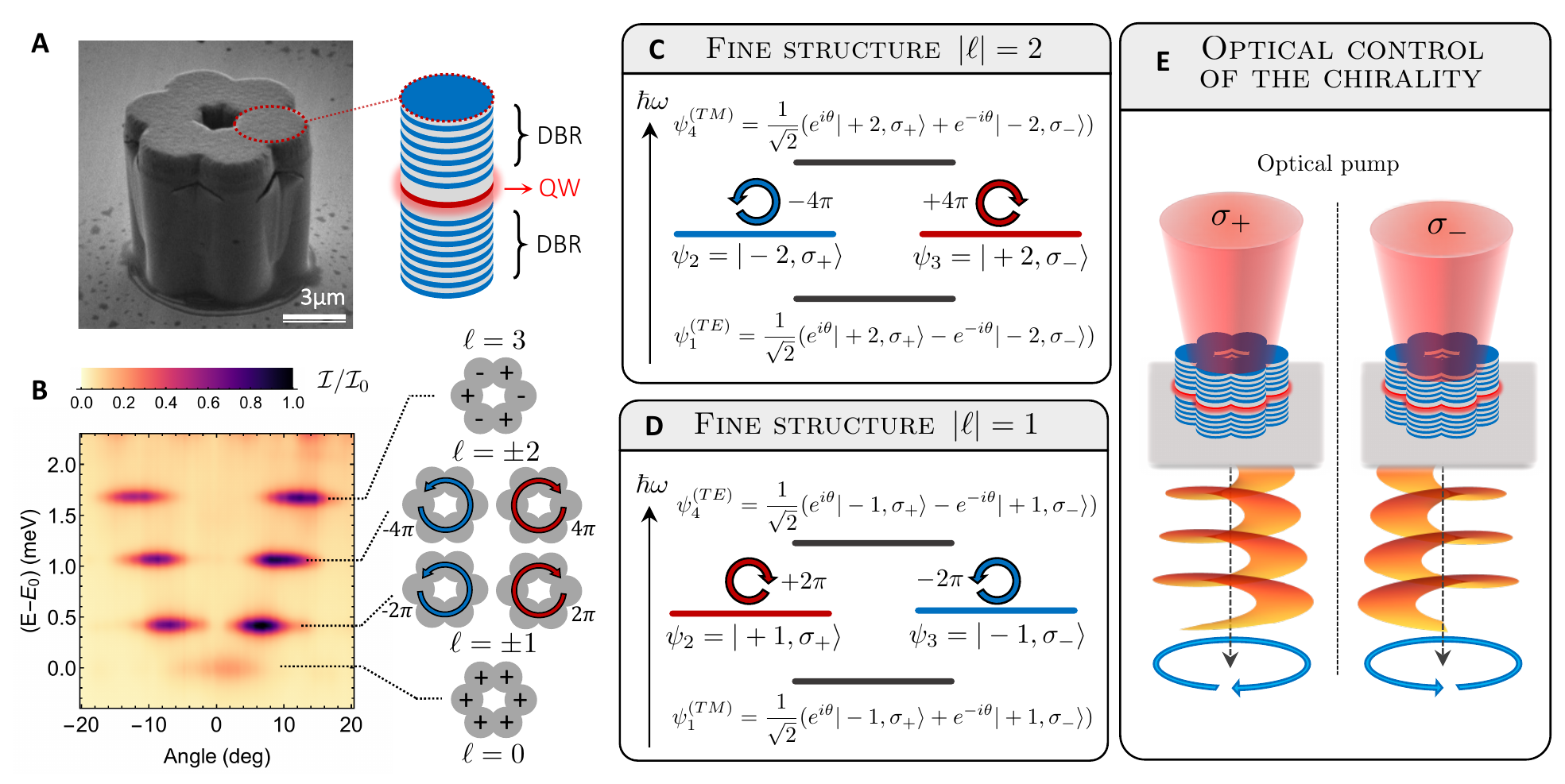}
		\caption{\textbf{Spin-orbit coupling in benzene-like OAM lasers.} (A) Scanning electron microscopy image of the benzene photonic molecule. On the right, a schematic representation of a single pillar embedding a quantum well between distributed Bragg mirrors (DBR) is sketched. (B) Angle-resolved photoluminescence spectrum below the lasing threshold present the four OAM manifolds ($\ell=0,\pm1,\pm2,3$). The face winding of each eigenmode (without accounting for the spin of photons) is schematically presented on the right. (C-D) Fine structure of the $\ell=2$ (C) and $\ell=1$ (D) manifolds when the spin-orbit coupling is taken in account. The red and blue lines correspond to states carrying a net CCW and CW OAM, respectively. In modes $\psi_{1,4}$, the phase $\theta=2\pi/6$. (E) Schematic representation of the protocol used in this work. A circularly polarized pump is used to trigger lasing in a benzene photonic molecule; the lasing emission carries a net OAM whose chirality is dictated solely by the polarization of the pump.}
		\label{structure}
	\end{figure*}
	
	\noindent\textbf{Spin-orbit coupling in benzene-like photonic molecules}
	
	To implement this scheme, we consider a vertical cavity surface emitting laser (VCSEL) formed from a semiconductor planar microcavity embedding a single $\mathrm{In_{0.05}Ga_{0.95}As}$ quantum well (details of the fabrication are presented in the Supplemental Materials); in such a cavity, lasing occurs in the weak coupling regime so that polariton is not involved above threshold. The cavity is then etched to form hexagonal rings of coupled micropillars (a scanning electron microscopy image of a ring is presented in Fig. \ref{structure} A). Due to the hybridization of the eigenmodes of each pillar, these benzene-like photonic molecules present six eigenmodes (without considering the spin of photons) that can be classified by their angular momentum $\ell$ associated to the relative phase between the pillars:
	
	\begin{equation}
		\label{mode}
		\ket{\ell}=\frac{1}{\sqrt{6}}\sum\limits_{j}e^{2\pi i\ell j/6}\ket{\phi_{j}},
	\end{equation}
	
	\noindent where $\ket{\phi_{j}}$ is the photonic ground state of the $j^{th}$ pillar. The four energy levels ($\ell\mathrm{=0,\pm1,\pm2,3}$) can be observed by angle-resolved non-resonant photoluminescence measurements below the lasing threshold (see Fig. \ref{structure} B). As schematically depicted in the right part of Fig. \ref{structure} B, the mode $\ell\mathrm{=0}$ presents a constant phase around the molecule and $\ell\mathrm{=3}$ a $\pi$ phase shift between neighbouring pillars; therefore, these states do not exhibit a net angular momentum. On the contrary, $\ell\mathrm{=\pm1}$ and $\mathrm{\pm2}$ present phase vortices with a topological charge $\ell$. 
	
	It was recently shown that, in these benzene-like photonic molecules, the spin moment of photons couples with its orbital angular momentum\cite{Sala2015}. This analogous spin-orbit coupling arises from the fact that photon hopping between neighbouring pillars is polarization-dependent: the coupling is typically $\mathrm{5-10~\%}$ greater for photons linearly polarized along the axis linking the pillars than for photons polarized perpendicularly\cite{MichaelisdeVasconcellos2011}. Under this effect, $\ell$ is no longer a good quantum number and the degeneracy of the $|\ell|=1$ and $|\ell|=2$ manifolds is lifted, each of them presenting a 3-level fine structure presented schematically in Fig. \ref{structure} C and D respectively. This fine structure can not be probed below the lasing threshold (e.g. in Fig. \ref{structure} B), because the linewdiths exceed the energy splittings; in the lasing regime, however, the linewidths become much smaller than the splittings allowing to observe emission from one single state of the manifold \cite{Sala2015}.
	
	\begin{figure*}
		\includegraphics[trim=0cm 0cm 0cm 0cm, width=150mm]{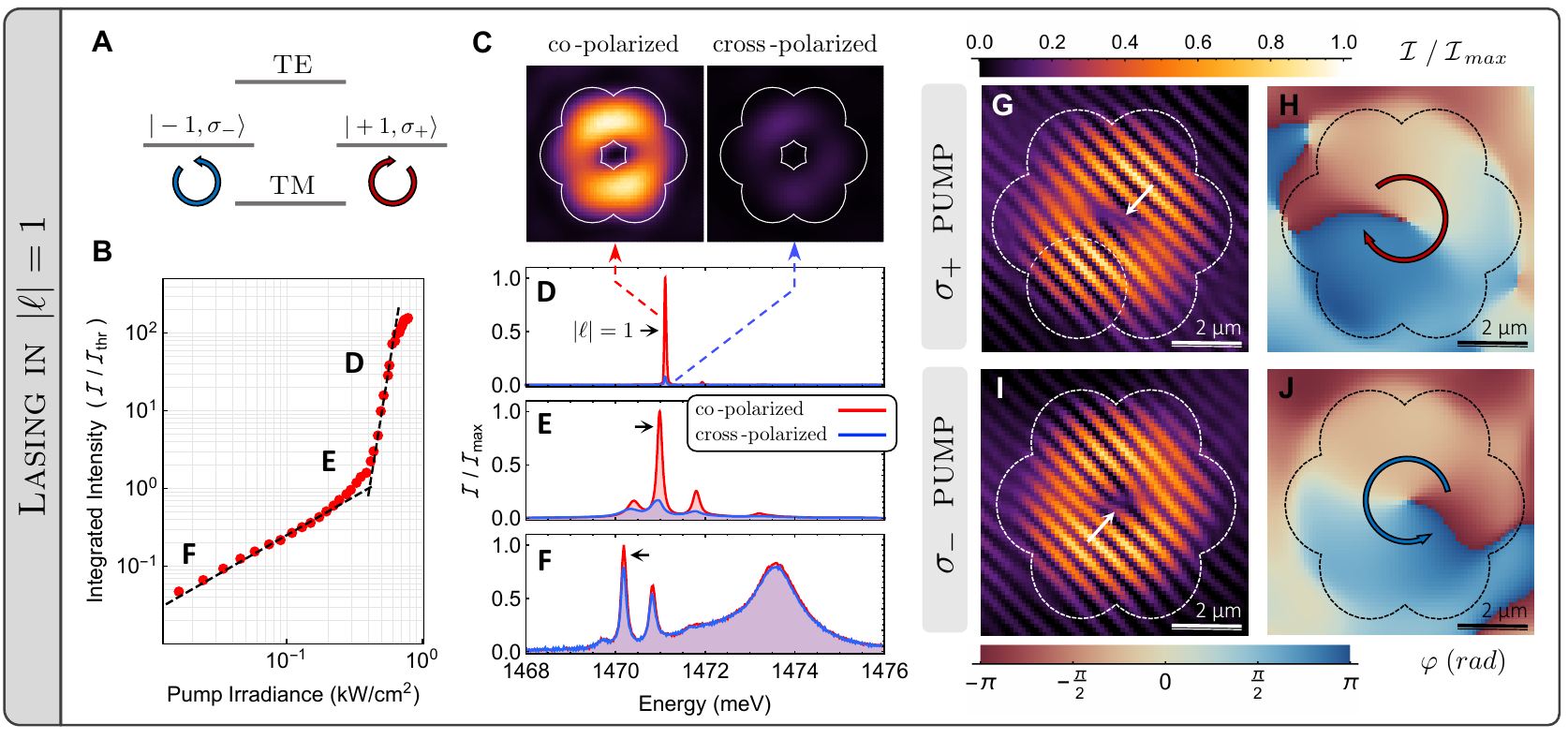}
		\caption{
			\textbf{Orbital angular momentum lasing in $|\ell|=1$ manifold}. (A) Fine structure of the $|\ell|=1$ manifold. (B) Integrated output intensity measured as a function of incident pump power showing a lasing threshold at $\mathrm{P_{thr}\sim 0.4~kW/cm^{2}}$. (C) Co- and cross-polarized real space images of the emission under a $\sigma_{+}$ polarized pump without spectral filtering. (D-F) Polarization and energy resolved emission spectra under $\sigma_{+}$ polarized pump above (D), at (E) and below (F) the lasing threshold. Red and blue curves correspond to co-($\sigma_{+}$) and cross-polarized ($\sigma_{-}$) emission with respect to the polarization of the pump.(G-I) Interference patterns measured as described in the text at $\mathrm{P=3~mW}$ (corresponding to Panel D) for a $\sigma_{+}$ and $\sigma_{-}$ polarized pump, respectively. (H-J) Corresponding phase maps showing $2\pi$ CW (CCW) vortex under $\sigma_{+}$ ($\sigma_{-}$) pump.}
		\label{LasingPanel}
	\end{figure*}
	\begin{figure*}
		\includegraphics[trim=0cm 0cm 0cm 0cm, width=150mm]{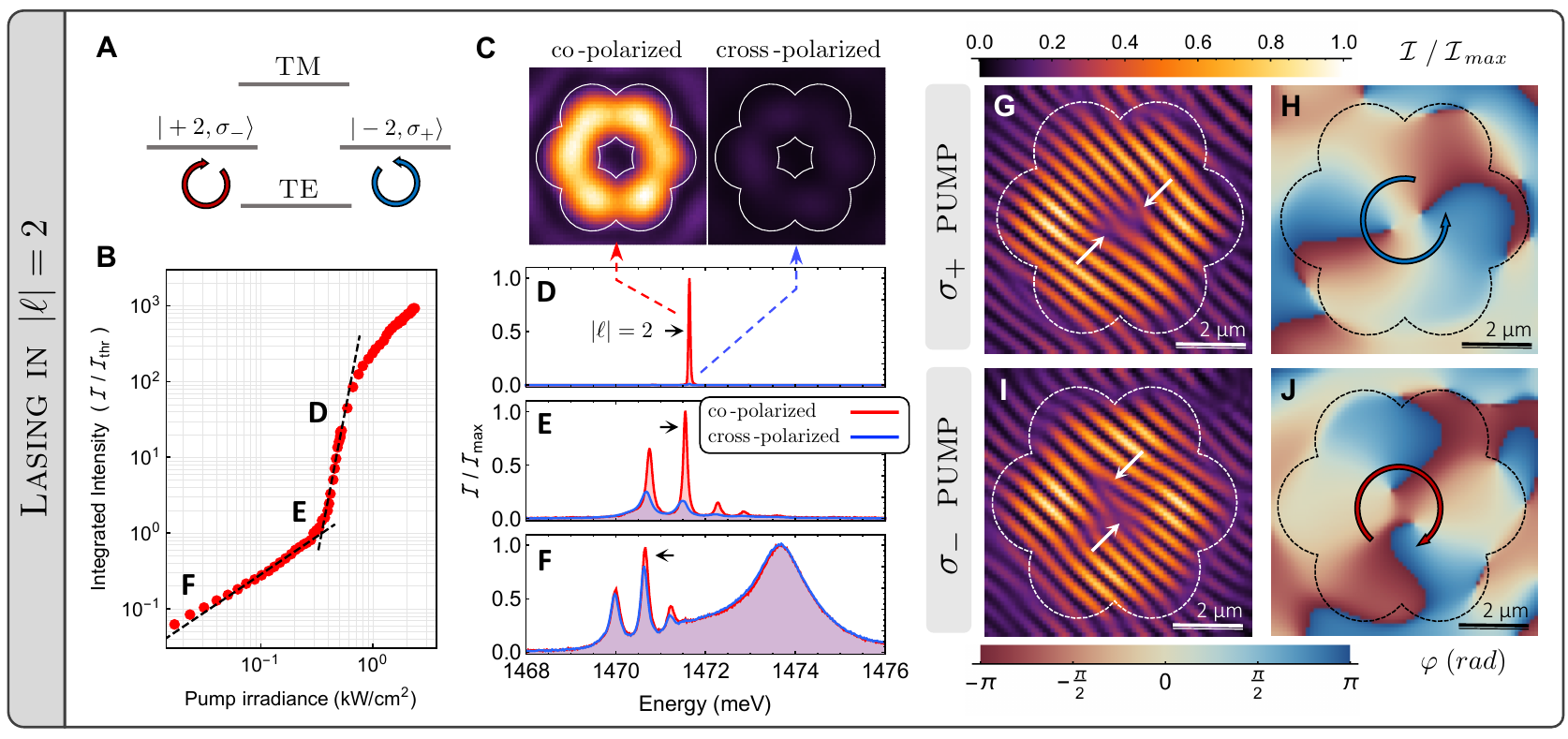}
		\caption{
			\textbf{Orbital angular momentum lasing in $|\ell|=2$ manifold}. (A) Fine structure of the $|\ell|=2$ manifold. (B) Integrated emission intensity measured as a function of incident pump power showing a lasing threshold at $\mathrm{P_{thr}\sim 2~kW/cm^{2}}$. (C) Co- and cross-polarized real space images of the emission under a $\sigma_{+}$ polarized pump. (D-F) Polarization and energy resolved emission spectra under $\sigma_{+}$ polarized pump above (D), at (E) and below (F) the lasing threshold. Red and blue curves correspond to co-($\sigma_{+}$) and cross-polarized ($\sigma_{-}$) emission with respect to the polarization of the pump.(G-I) Interference patterns measured as described in the text at $\mathrm{P=0.6~kW/cm^{2}}$ (corresponding to Panel D) for a $\sigma_{+}$ and $\sigma_{-}$ polarized pump, respectively. (H-J) Corresponding phase maps showing $4\pi$ CCW (CW) vortex under $\sigma_{+}$ ($\sigma_{-}$) pump.}
		\label{LasingPanel2}
	\end{figure*}
		
	In this fine structure, the highest- and lowest-energy modes ($\psi_{1,4}$) present either a radial (TE) or azymuthal (TM) linear polarization, and do not exhibit orbital angular momentum. They are formed from a linear combination of CW and CCW vortices (see wave-functions in Fig. \ref{structure} C and D, and see Supplemental Materials for their derivation). On the other hand, the degenerate middle modes present opposite circular polarizations ($\psi_{2,3}$), and carry net OAM with opposite chiralities. To obtain lasing in a chiral mode, one needs to pump preferentially one of these two degenerate states. The major asset of the fine structure in our device is that for a spin-polarized gain medium, one of these chiral modes presents the highest gain of the manifold (either $\psi_{2}$ or $\psi_{3}$ depending on the polarization): if the medium is fully polarized, this gain is twice larger than for the radially or azimuthally polarized modes $\psi_{1}$ and $\psi_{4}$. This allows to trigger lasing in either of these chiral modes by injecting spin-polarized carriers in the device. As presented hereafter, this can be achieved with a circularly polarized off-resonant optical pump (as schematically presented in Fig. \ref{structure} E); therefore, the chirality of the emission can be controlled by simply tuning the polarization of the pump.\\

	\noindent\textbf{OAM lasing with optically controlled chirality}
	
	To demonstrate this optical control of the chirality, we investigated two different devices formed from photonic molecules with $\mathrm{3.2~\mu m}$-diameter micropillars, and an inter-pillar distance of $\mathrm{2.3~\mu m}$ (molecule M1) or $\mathrm{2.4~\mu m}$ (molecule M2). The variation of the inter-pillar distance modifies the relative gain/loss ratio of the photonic modes (see Supplemental materials), allowing to select a precise $|\ell|$ manifold in which lasing occurs: for the molecule M1 (M2), lasing occurs in the $|\ell|=1$ ($|\ell|=2$) manifold.
	
	The devices were held in a closed cycle cryostat at $\mathrm{T=4~K}$. Fig. \ref{LasingPanel} presents the results for molecule M1 when exciting the device with a circularly polarized ($\sigma_{+}$) off-resonant pump ($\mathrm{E_{pump}\sim1.6~meV}$). A nonlinear increase of the integrated emission intensity is observed above a threshold power density of $\mathrm{P_{th}=0.4~kW/cm^{2}}$ (see panel B); emission spectra for excitation powers above, around and below this threshold are respectively presented in panels D, E and F. Above the threshold, we see the emergence of a single mode emission from the $|\ell|=1$ manifold, as well as a strong narrowing of the linewidth, being only limited by the resolution of the spectrometer ($\mathrm{\sim 40~\mu eV}$). This unambiguously indicates the onset of lasing in this manifold.
	
	At low excitation power (Fig. \ref{LasingPanel} F), the emission presents a non-negligible degree of polarization ($\mathrm{P=\frac{\mathcal{I}\sigma_{+}- \mathcal{I}\sigma_{-}}{\mathcal{I}\sigma_{+} + \mathcal{I}\sigma_{-}}\sim5~\%}$), demonstrating that the spin polarization of photo-generated carriers (imposed by conservation of angular momentum of the circularly polarized pump photons) is significantly preserved during their relaxation in the gain medium (i.e. the quantum well). In the lasing regime (Fig. \ref{LasingPanel} D-F), $P$ is greatly enhanced reaching almost unity ($\mathrm{P>95~\%}$), thanks to the stimulated nature of the emission. Fig. \ref{LasingPanel} C presents real space images of the device emission (without any spectral filtering) under co- and cross-polarized detection evidencing that the whole beam presents this strong degree of polarization. This indicates that the spin-polarized pump indeed triggers lasing in a circularly polarized mode.
		
	In order to evidence the phase vortex associated to this lasing mode, we interfere the beam with a magnified image of the emission from one of the molecule pillars, which acts as a phase reference. The resulting interferogram, taken without spectral nor circular polarization filtering, is shown in Fig. \ref{LasingPanel} G (the dashed circle indicates the reference pillar).  The corresponding phase map, obtained with a standard off-diagonal Fourier filtering technique, is presented in Fig. \ref{LasingPanel} H. The pitchfork in the interferogram (marked by a white arrow) and the vortex in the phase map evidence a $\mathrm{2\pi}$ winding of the phase around the molecule, showing that the laser mode presents an OAM of $\ell=+1$ (corresponding to $\ket{\ell=+1,\sigma_{+}}$ in Fig. \ref{LasingPanel} A). Remarkably, when changing the excitation polarization to $\sigma_{-}$, the polarization of the gain medium is inverted, and the lasing mode, fully $\sigma_{-}$ polarized, now evidences opposite chirality, corresponding to $\ell=-1$. This is evidenced by the interferogram and corresponding phase map presented in Fig. \ref{LasingPanel} I and J that respectively show an inversion of the pitchfork and of the circulation of the phase. This demonstrates the ability to optically control the chirality of the lasing mode.
			
	\begin{figure}
		\includegraphics[trim=0cm 0cm 0cm 0cm, width=75mm]{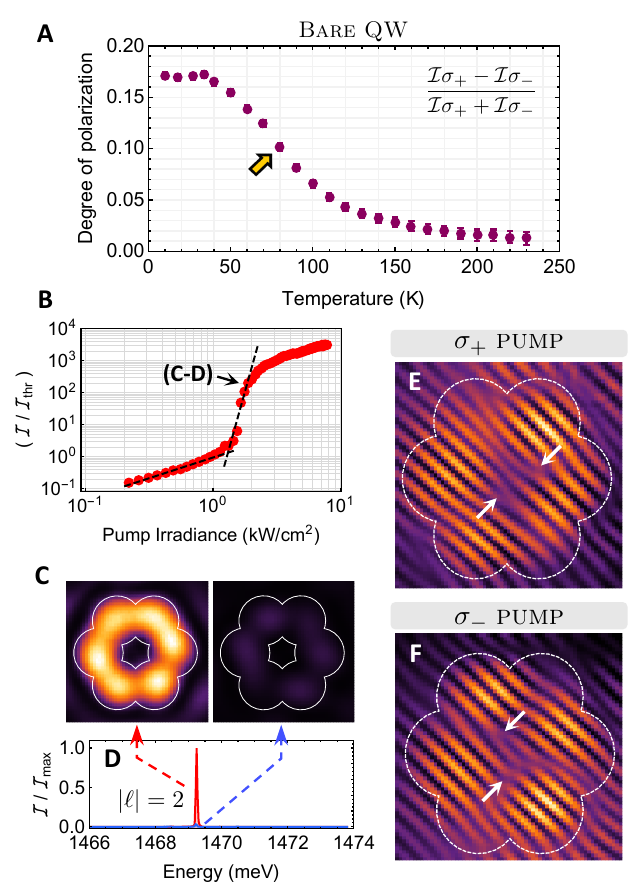}
		\caption{
			\textbf{Operation at 80K}. (A) Degree of circular polarization of the bare quantum well photoluminescence as a function of temperature. (B) Integrated intensity of Device M2 as a function of the pumping power. (C) Real space images of the beam and (D) emission spectra, in the lasing regime (at the energy indicated in Panel B), for a detection co- and cross polarized with respect to the $\sigma_{+}$ pump. (E-F) Self-interference of the laser emission showing the characteristic $\pm 4\pi$ phase vortices under a $\sigma_{+}$ (E) and $\sigma_{-}$ (F) pump.} 
		\label{OAM80KPanel}
	\end{figure}
	
	When considering the second device (molecule M2), we now observe that a $\sigma_{+}$ polarized pump triggers lasing in the $|\ell|=2$ manifold (see Fig. \ref{LasingPanel2} B for the I-P curve, and D-F for the emission spectra). Similarly as for M1, the degree of polarization of the emission in the lasing regime is very strong ($\mathrm{P>95\%}$) and dictated by the polarization of the pump. The phase vortex of this lasing beam is extracted using the same technique as for M1: the interferogram and corresponding phase map are presented in Fig. \ref{LasingPanel2} G and H, respectively. The two pitchforks in the interferogram (white arrows) and the double phase vortex clearly show that the emission now presents an OAM of $\ell=-2$. Then, by changing the polarization of the pump to $\sigma_{-}$, the chirality of the emission is inverted to $\ell=+2$, as show the interferogram and phase map presented in \ref{LasingPanel2} I and J.\\

	\noindent\textbf{Temperature robustness of the scheme}
		
	The possibility to control the chirality of the emission only requires that the spin-dependence of the gain within one OAM manifold dominates all other possible contributions (e.g. spectral dependence of the gain). In our case, this condition is fulfilled by optimizing the gain at the energy of a precise $|\ell|$ manifold and imprinting a sufficiently large spin polarization in the quantum well. This latter condition can be undermined by the onset of thermally activated spin-relaxation processes. In order to evaluate the robustness of our devices against these processes, we measured the degree of circular polarization of the photoluminescence from a single quantum well (without the cavity) as a function of temperature, when pumped with a $\sigma_{+}$ polarized pump (see Fig.\ref{OAM80KPanel} A). At $\mathrm{T=4~K}$, the emission presents a degree of circular polarization of $\mathrm{\sim 17~\%}$. Interestingly, the degree of polarization remains non-negligible ($\mathrm{\sim10~\%}$) up to $\mathrm{T=80~K}$; at higher temperature ($\mathrm{T>100~K}$), it slowly vanishes below $\mathrm{5~\%}$. 
	
	As a result of this temperature resilience, we were able to implement our scheme up to $\mathrm{T=80~K}$: Fig.\ref{OAM80KPanel} B-F shows that exciting device M2 with circular polarization still allows triggering lasing with an optically controllable OAM $\ell=\pm2$. Above this temperature, the emission of the laser was no more circularly polarized, indicating the loss of the carrier polarization during the thermalization of the carriers. Consequently, the lasing mode did not exhibit any net chirality.\\
	
	\noindent\textbf{Perspectives}
	
	It is important to point out that the scheme we demonstrate here is not restricted to benzene-like molecules, but can be implemented in any $n$-pillars ring molecules with $n$ even and $\ge 4$ (see Supplemental Materials for a demonstration based on symmetry group arguments). When considering $n$ pillars, the $|\ell|=1$ and $|\ell|=n/2-1$ manifolds present the adequate fine structure similar to that of $|\ell|=1$ and $|\ell|=2$ in Fig. \ref{structure} C and D. This could pave the way to the implementation of microlasers generating arbitrarily large OAM with tunable chirality. As well, the demonstrated scheme allows, in principle, controlling the emission chirality on ultrashort timescales, limited only by the relaxation times of carriers ($\mathrm{\sim ps}$). Both of these aspects are highly relevant for quantum and classical information transmission protocols.
	
	Furthermore, our design can be transposed to other laser architectures, because the underlying concepts (i.e. spin-orbit coupling of photons and spin-polarization of the gain medium) are very general. For instance, active materials with more robust spin properties, such as transition metal dichalcogenide monolayers (e.g. MoS2 and WSe2) which present electron spin-valley locking, could allow reaching room temperature operation\cite{Wang2018a}. Also, combining the present design with ferromagnetic electrodes\cite{Allain2015, Wolf2001} could open the possibility to fabricate OAM microlasers with electrical injection.
	
	Finally, our results could be extended beyond the field of OAM lasers, for example in optomechanics by coherently transferring photonic angular momentum to chiral torsional modes of the microstructures\cite{Fenton2018}. As well, by embedding a quantum emitter (e.g. a quantum dot or semiconductor defects) in the resonator instead of a quantum well, it would be possible to generate single photons with a controllable OAM\cite{Fong2018}.\\
	
	\noindent\textbf{Acknowledgements}
	\noindent The authors are thankful to Luis Orozco for fruitful discussions. This work was supported by the ERC grant Honeypol, by the EU-FET Proactive grant AQuS, the French National Research Agency (ANR) project Quantum Fluids of Light (ANR-16-CE30-0021) and the Labex CEMPI (ANR-11-LABX-0007) and NanoSaclay (ICQOQS, Grant No. ANR-10-LABX-0035), the French RENATECH network, the CPER Photonics for Society P4S, and the M\'etropole Europ\'eenne de Lille via the project TFlight. P. S.-J. acknowledges financial support from the Marie Curie individual fellowship ToPol and from the Natural Sciences and Engineering Research Council of Canada (NSERC).

	\clearpage

	\onecolumngrid

	\renewcommand{\thefigure}{S\arabic{figure}}
	
	\begin{center}
		\huge{\textbf{\textsc{Supplementary material}}}
	\end{center}
	\section{Sample description}
	
\begin{figure*}[h]
		\centering
		\includegraphics[scale=1.0]{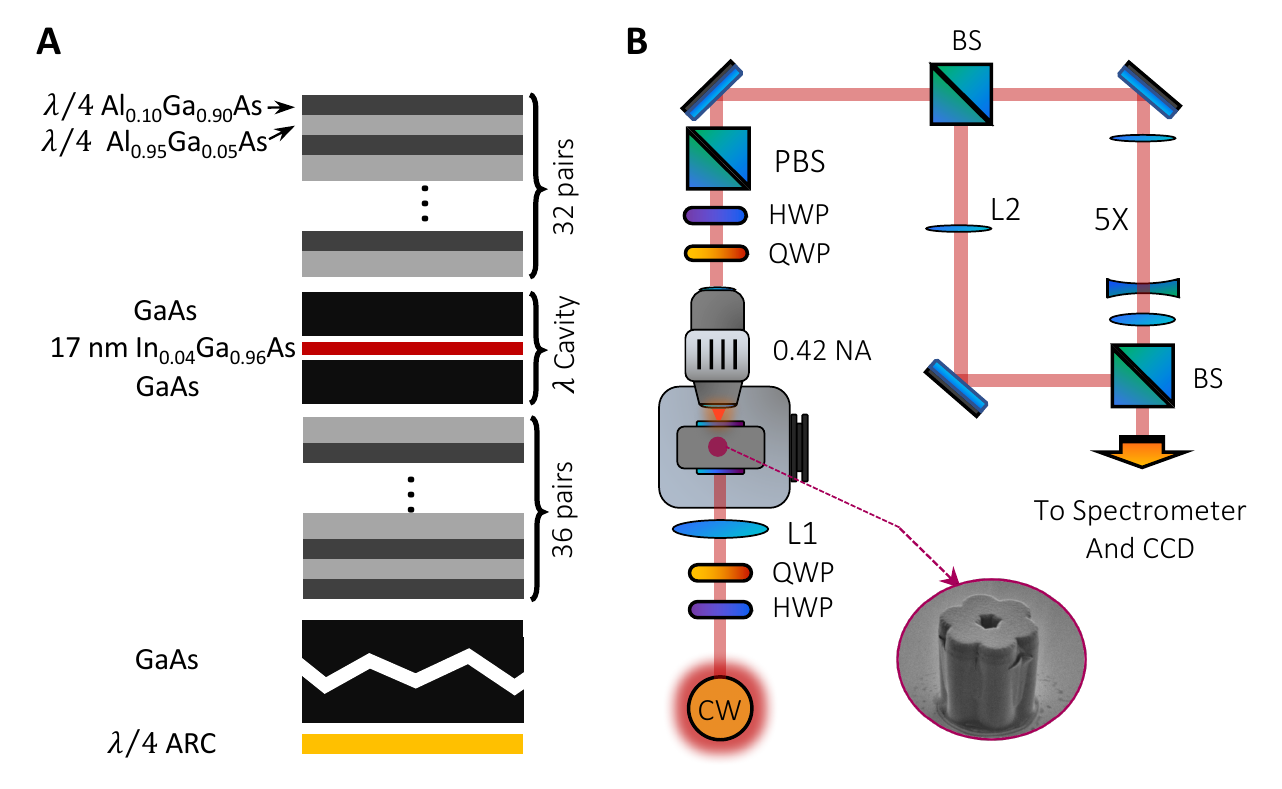}
		\caption{\textbf{Sample and setup details}: (A) Cross-section of the semiconductor heterostructure forming the laser cavity. (B) Sketch of the experimental setup. The inset shows a scanning electron microscopy image of a device.}  
		\label{SampleAndSetup} 
	\end{figure*}

	The benzene microstructures were fabricated from a planar semiconductor heterostructure grown by molecular beam epitaxy, consisting of a $\lambda$ Fabry-Pérot cavity embedding a 17~nm wide $\text{In}_{0.04}\text{Ga}_{0.96}\text{As}$ quantum well. The cavity is formed by a GaAs spacer enclosed between two $\text{Al}_{0.10}\text{Ga}_{0.90}\text{As} \, / \, \text{Al}_{0.95}\text{Ga}_{0.05}\text{As}$ distributed Bragg reflectors (DBR), with 32 and 36 pairs of layers at the top and bottom, respectively. The cavity spacer presents a thickness gradient, allowing to tune the cavity mode with respect to the gain spectral dependence. The measured quality factor of the planar cavity is $Q\approx 4\cdot10^4$, limited mainly by residual absorption in the spacer. The cavity was grown on a double-side polished 350~$\mu$m thick GaAs substrate, in order to allow operation in transmission geometry. The planar structure was then dry-etched to form coupled micropillars
	structures with different pillar diameters and inter-pillar distances. Finally, to prevent multiple reflections at the interface between GaAs and vacuum we deposited an anti-reflection coating (ARC), consisting of a silicon oxynitride quarter wavelength layer. An overall view of the planar structure is sketched in Fig. \ref{SampleAndSetup} A. 
	
	\section{Setup and Measurements}
	
	A sketch of the experimental setup used in this work is presented in Fig. \ref{SampleAndSetup} B. The sample was held in a closed-cycle cryostat where the temperature could be actively stabilized between 4K and room temperature. The device was pumped on the epitaxial side (top in Fig. \ref{SampleAndSetup} A) with a continuous wave Ti:Sapphire laser focused by a lens with a focal length of 100~mm producing a 20~$\mu$m FWHM gaussian spot, ensuring a quasi-uniform illumination of the structure. The incident polarization state of the pump was controlled with a $\lambda/4$ waveplate, and its wavelength was set to $770$~nm corresponding to one of the reflectivity minima of the DBRs.
	
	The emission from the microstructure was collected from the substrate (bottom in Fig. \ref{SampleAndSetup} A) side using a $0.42$~NA objective and its polarization state was analyzed with a polarizing beam splitter (PBS) together with a quarter-wave plate (QWP) and a half-wave plate (HWP) allowing to realize a full polarization tomography. Energy resolved measurements (e.g. in Fig. 1 B, 2 D-F, 3 D-F and 4 D of the main text), were realized with a CCD camera coupled to a spectrometer; for measurements without spectral resolution (e.g. Fig. 2 C,G-J, 3 C,G-J and 4 C,E,F of the main text), we used the same setup but analyzed the detection at the $\mathrm{0^{th}}$ order of the spectrometer.
	
\begin{figure}[h]
		\centering
		\includegraphics[scale=0.8]{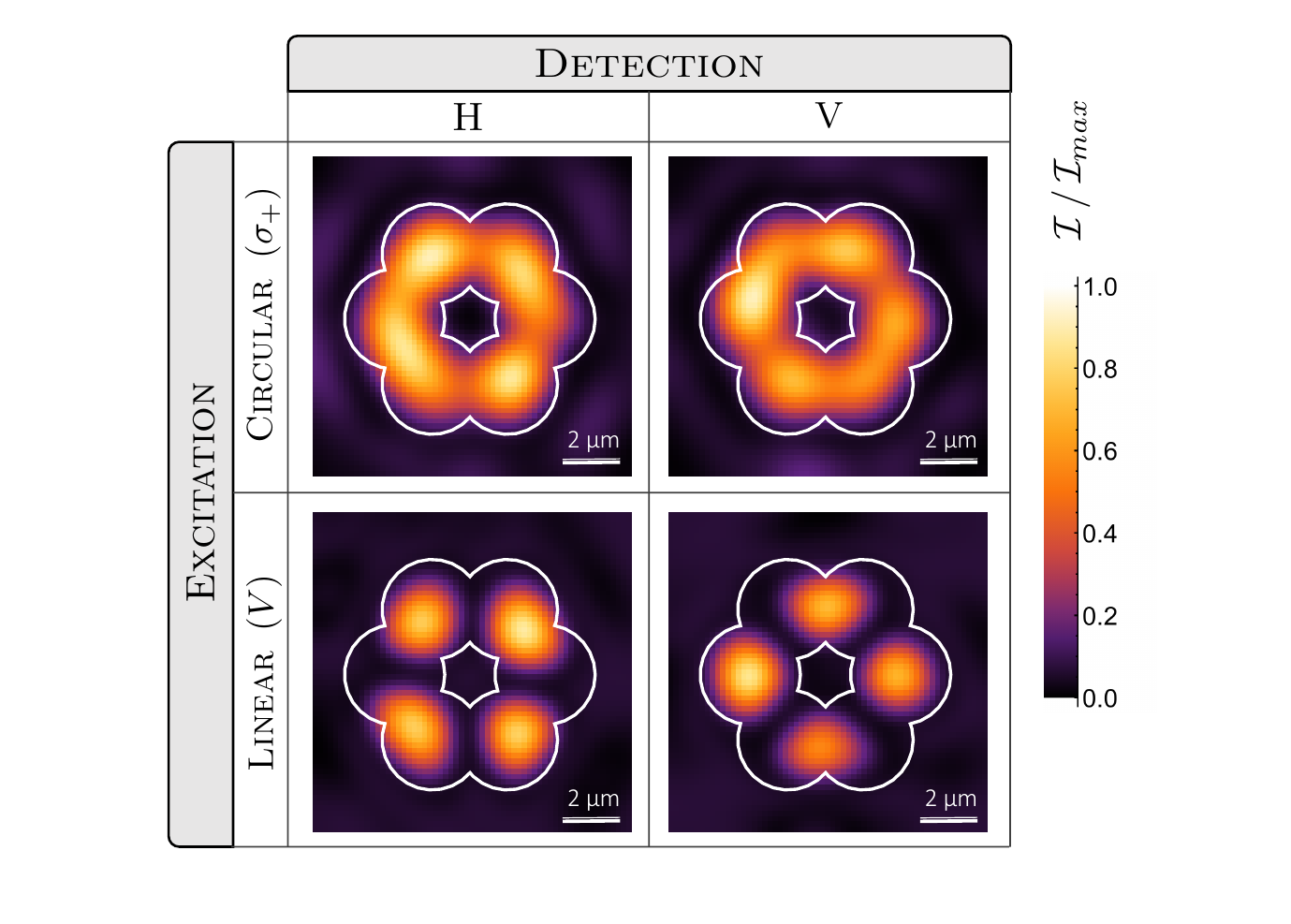}
		\caption{\textbf{Emission under linearly and circularly polarized excitation.} Real space images of the emission (without spectral filtering) for circularly polarized (A-B) and linearly polarized (C-D) excitations. All images are taken under a linear polarization: horizontal for (A) and (C) and vertical for (B) and (D).}  
		\label{LinCircExc}
	\end{figure}

	For the measurements that were realized without circular polarization filtering, we used only a polarizing beam splitter (PBS) in the detection path to select a well-defined linear polarization component of the emission. This was done in order to get rid of the effects of the polarization-dependent reflectivity of the spectrometer grating by sending always the same linear polarization in it. It is important to point out that this selection of a single linear polarization in the detection path does not affect the emission pattern of the modes that are circularly-polarized (i.e. $\psi_{2,3}$ which are the important modes for realizing our OAM lasing scheme), whereas it results in the formation of dark regions for modes which have a linearly-polarized emission pattern (i.e. $\psi_{1,4}$ which are radially and azymuthally polarized). This can be seen in Fig. \ref{LinCircExc} where we present real space images of the beam for circularly and linearly polarized excitation for a molecule with inter-pillar distance of $\mathrm{2.4~\mu m}$ (where lasing occurs in the $|\ell|=2$ manifold). The linearly polarized excitation triggers lasing in $\psi_{4}$, presenting a linear polarization; the circularly polarized one ($\sigma_{+}$ in this case) triggers lasing in $\psi_{3}$ which is unaffected by the linear polarization detection (H corresponds to the linear polarization selected by the PBS for all measurements without circular polarization selection).
	
	To realize the interferometry measurements, the optical path was separated in a modified Mach-Zender interferometer: in one arm (signal), a lens with a $750$~mm focal length imaged the real space emission on the spectrometer entrance, and, in the second arm (reference), an expanded image of the emission was created using a 5$\times$ magnifying telescope. When recombining the two images, we can overlap the emission of the whole microstructure with a magnified image of one of the pillars. It was possible to consider the magnified beam as a phase reference, because the phase gradient within the area of a single pillar is much smaller than $\pi$. Therefore, by slightly tilting the direction of the reference beam with respect to that of the signal beam after their recombination, we could produce fringe patterns as shown in the main text (e.g. in Fig. 2,3 G,I and 4 E,F).

	\section{Retrieving the phase maps}
	
\begin{figure}[hbtp]
		
		\centering
		\includegraphics[scale=0.9]{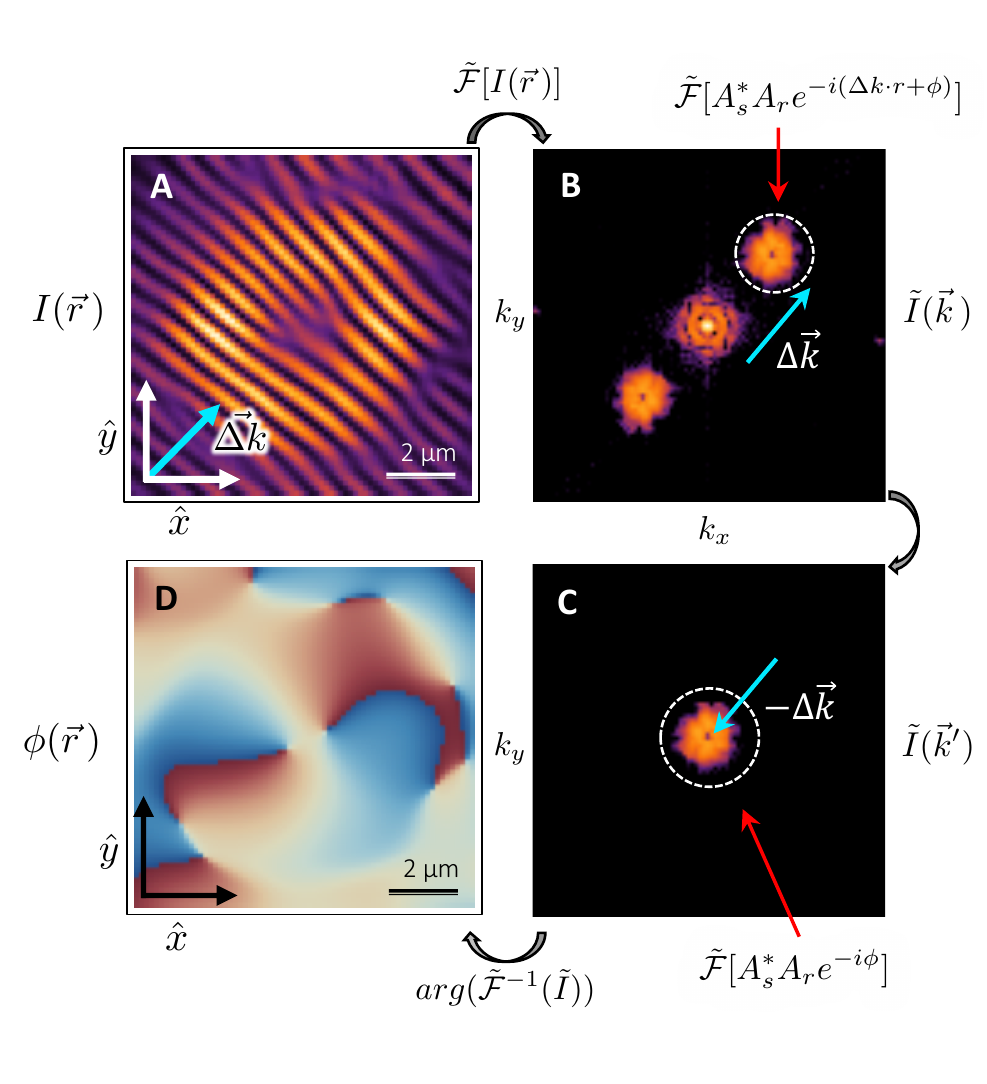}
		\caption{\textbf{Retrieving the wavefront phase}: (A)-Typical inteferrence pattern for the $\vert -2,\sigma_{+}\rangle$ mode (from Fig.3-G in the main text), produced by interfering the emission and reference arm with an in-plane wavevector difference $\Delta\mathbf{k}$. (B) Absolute value of the digitized interferogram fast Fourier transform (FFT). (C) After a rigid translation of the  FFT by $-\Delta\mathbf{k}$ we apply a radial low-pass filter. (D) Real space image of the emission phase $\phi(r)$, obtained as the argument of the inverse FFT.}
		\label{OffDiagFFT}
	\end{figure}

	In the main text we present in Fig. 2 and 3 interferograms for the $\ell=\pm1$ and $\ell=\pm2$ modes and, next to them, the corresponding phase maps. In order to extract these maps we used an off-diagonal Fourier filtering technique which we schematically depict in Fig. \ref{OffDiagFFT}. The idea is that if the signal and reference beams are described by the complex amplitudes:
	\begin{equation}
		\begin{aligned}
			\mathcal{A}_{s} &=  A_{s}(\mathbf{r}) e^{-i(\omega t -\mathbf{k}_s \cdot \mathbf{r} +\phi)}\\
			\mathcal{A}_{r} &= A_{r}(\mathbf{r})e^{-i(\omega t - \mathbf{k}_r \cdot \mathbf{r} )} ,
		\end{aligned}
	\end{equation}
	
	\noindent then the fringe intensity pattern produced at the spectrometer entrance plane ($z=0$) will be given by:
	\begin{equation}\label{eq:FringeIntensity}
		\begin{aligned}
			I(\mathbf{r})&=\vert \mathcal{A}_{s}  + \mathcal{A}_{r} \vert^2 \\
			&=\vert A_{s}\vert^2 + \vert A_{r}\vert^2 + \left( A_{s} A_{r}^{*} e^{-i(\Delta \mathbf{k}\cdot \mathbf{r} +\phi)} + c.c. \right),
		\end{aligned}
	\end{equation}
	
	\noindent where $\Delta \mathbf{k}$ is the in-plane wavevector difference between the signal and reference beams. One example of these fringe patterns is provided in Fig. \ref{OffDiagFFT} A. If we Fourier-transform the fringe pattern $\tilde{I}( \mathbf{k})=\mathcal{\tilde{F}}[I]$, the first two terms in equation \eqref{eq:FringeIntensity} produce a peak centered at the reciprocal space origin, whereas the third and fourth term correspond to two satellite peaks translated by $\pm\Delta \mathbf{k}$ with respect to the origin. In Fig. \ref{OffDiagFFT} B we show the modulus of the Fourier-transformed fringe pattern. These satellite peaks carry the information on the wavefront phase, which can be retrieved by operating a rigid translation of the reciprocal space by $\pm\Delta \mathbf{k}$ and filtering-out all the other peaks (see Fig. \ref{OffDiagFFT} C). The remaining peak corresponds to $\mathcal{\tilde{F}}[A_{s} A_{r}^{*}e^{-i\phi}]$. Then, by inverse Fourier transforming this complex amplitude and taking its argument, we can extract the phase pattern (Fig. \ref{OffDiagFFT} D) associated to the initial fringe pattern (Fig. \ref{OffDiagFFT}-A). Remarkably, a clear signature of the phase singularity present in the fringe pattern, can be directly observed also in reciprocal space, where it manifests as a dark spot in the middle of the side-peaks.
	
	\begin{figure}[h]
		\centering
		\includegraphics[scale=1.0]{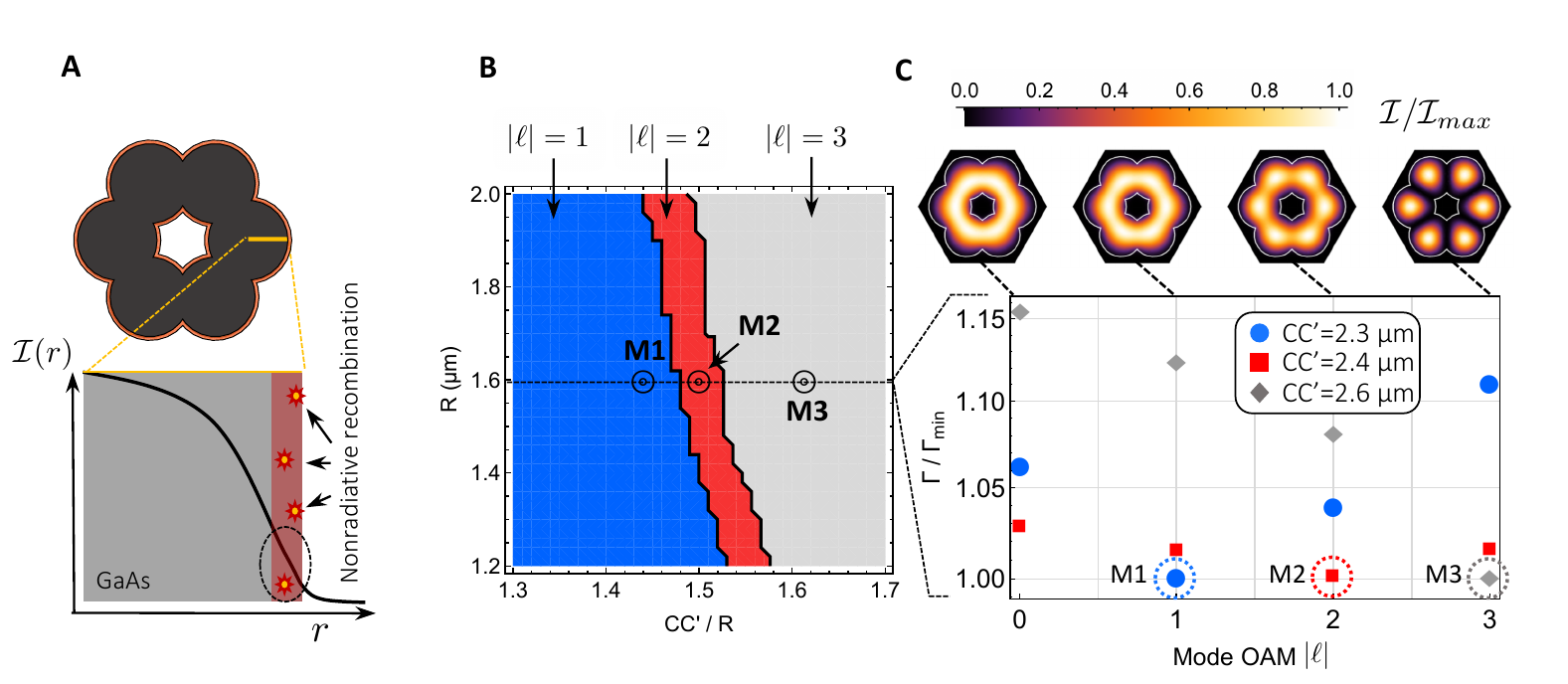}
		\caption{\textbf{Lasing mode selection}: (A) Schematic representation of the device; the orange region along the perimeter highlights the region where the density of nonradiative recombination centers is the highest. The zoom-in shows an image of the lateral section of one pillar near the outer edge, and the solid line represents a typical radial profile of a mode. (B) Mapping presenting which manifold is expected to exhibit the lowest non-radiative losses as a function of the inter-pillar distance and pillar radius. The blue, red and gray areas correspond to regions where, respectively, $|\ell=1,2,3|$ manifold exhibits the lowest losses. The black circles correspond to the three different devices (M1, M2 and M3) investigated in this work. (C) Plot of the calculated non-radiative losses associated to each scalar eigenmode, where $\Gamma_{min}$ corresponds to the lowest value calculated. Blue circles, red squares and grey diamonds correspond to structures with pillar radius of $\mathrm{R=1.6~\mu m}$ and an inter-pillar distance of, respectively, $CC'=(2.3,\, 2.4,\, 2.6)\mu$m. On top we show the calculated scalar eigenmodes intensity profile for $CC'=2.4\mu$m.}
		\label{OAMselectionLosses}
	\end{figure}
	
	\section{Selection of the lasing mode}
	
	In the main text, we show that two microstructures, labeled M1 and M2, having the same radius ($1.6~\mathrm{\mu m}$) but different interpillar distance, lase in modes $\vert\ell\vert=1$ and $\vert\ell\vert=2$ respectively. In this section we provide an explanation based on a model with a minimal number of parameters describing how it is possible to tune the relative gain/loss ratio of each OAM manifold (and thus select in which of them lasing occurs) by changing this interpillar distance.
	
	Firstly, the scalar eigenmodes of the photonic molecule (i.e. $\vert\ell\vert=0,...,3$) span over only few meV, a small energy scale compared to the typical spectral width of the gain profile. This allows neglecting any significant spectral dependence of the gain. Since the structure was evenly illuminated, we can as well consider that the spatial overlap between the gain medium and every mode is nearly identical. 
	
	The radiative lifetime given by photons leaking through the mirrors is the same for each mode. However, the dry etching process induces near the pillar edges a high density of nonradiative recombination centers (NRC) which represent the main mode-dependent contribution to the gain/loss ratio.
	
	We can deduce the relative contribution of these nonradiative losses by computing the overlap integral of each eigenmode with the NRC density profile. For simplicity, we assume the latter to be uniform over a thickness $\delta R$ along the edges of the structure, as  sketched in Fig. \ref{OAMselectionLosses} A. The scalar photonic eigenmodes of the microstructure were obtained by finite element methods: the hexagonal molecules were described as six overlapping infinite waveguides with the same transverse profile as the microstructure, and we solved the associated Helmholtz equation.The refractive index of the waveguides was chosen to match the effective refractive index of the vertical cavity ($n_{\mathrm{eff}}\approx3.46$ at $\mathrm{10~K}$). This effective refractive index can be computed by weighting the refractive index profile $n(z)$ by the vertical field profile $u(z)$ obtained for a planar cavity with transfer matrix methods.
	
	According to these calculations, we show in Fig. \ref{OAMselectionLosses} B which mode has the lowest nonradiative contribution (and therefore should lase) as a function of the micropillar radius $R$ and inter-pillar distance $CC'$ (red, blue and gray areas correspond to $|\ell|=1$, $|\ell|=2$ and $\ell=3$ manifolds, respectively). We see that for sufficiently small inter-pillar distance the favoured manifold is $\vert\ell\vert=1$ whereas for large $CC'$ values the favoured mode is $\ell=3$; in-between these two regions, when $CC'\sim 3/2\, R$ , the $\vert\ell\vert=2$ mode is favoured. This behaviour stems from the competition between two different mechanisms. On the one hand, modes with increasing $\vert\ell\vert$ have a profile that peaks at larger values of the radial coordinate, thus having a greater overlap with the outer edge of the microstructure. On the other hand, modes with a smaller $|\ell|$ present a larger overlap with inter-pillar regions where the density of NRC is the largest; this second effect is more pronounced as $CC'$ increases.
	
	The exact range of inter-pillar distances $CC'/R$, as a function of $R$, where the $\vert\ell\vert=2$ mode is favoured, depends on the thickness of the nonradiative recombination region considered. In order to match the experimental observations (black circles), we had to set $\delta R\approx \mathrm{0.1~\mu m}$, which is a reasonable value for the etching technique used. The black circles correspond to the microlaser structures presented in Fig. 2 and 3 in the main manuscript (labeled as M1 and M2) and to a structure with $CC'=\mathrm{2.6~\mu m}$ (M3) where lasing occurs in the $\ell=3$ manifold (see Section V). In Fig. \ref{OAMselectionLosses} C, we plot the relative nonradiative linewidth $\Gamma/\Gamma_{min}$, where $\Gamma=\hbar/\tau_{NR}$ and $\Gamma_{min}$ corresponds to the lowest value calculated for $\Gamma$. On top of this panel, we present the typical intensity profile of each scalar eigenmode of the system, for microstructures associated to the black circles in Panel B (indicating the structures M1,M2 and M3). This elementary model provides a satisfactory understanding of the mechanism driving the mode selection in these microstructures.
	
	\section{Lasing in $\ell$=3 modes}
	
\begin{figure}[htbp]
		\centering
		\includegraphics[scale=0.85]{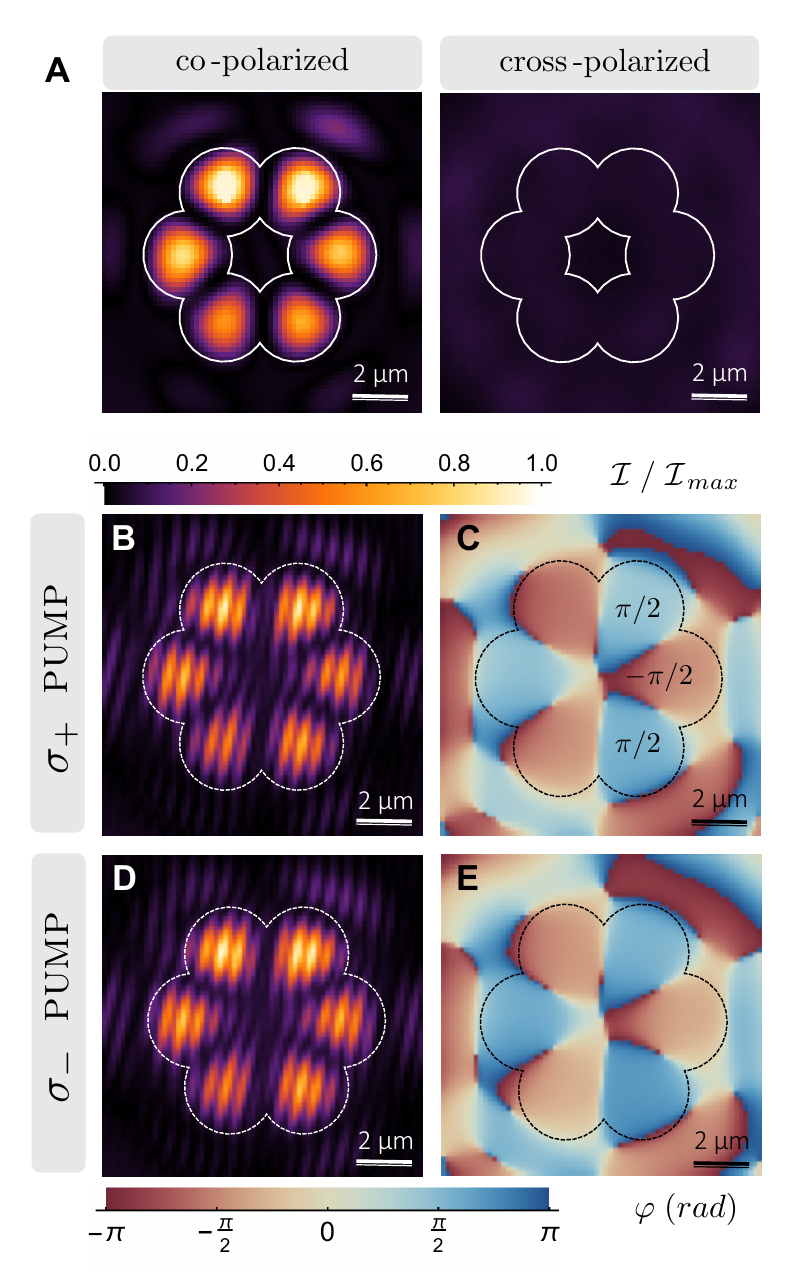}
		\caption{\textbf{Lasing in the $\vert\ell\vert=3$  manifold}: (A) Co- and cross-polarized real space images of the emission under $\sigma_{+}$ polarized pump. The inter-pillar distance here is $CC'=2.6\,\mu$m, the images were taken for an incident excitation density of $\approx0.7\,$kW/$cm^2$. (B,D) Interference pattern obtained as described above for a $\sigma_{+}$  and $\sigma_{-}$  polarized pump, respectively. (C,E) Corresponding phase maps showing the characteristic $\pi$ phase jumps between neighbouring pillars.}
		\label{LasingInL3}
	\end{figure}

	In figure \ref{LasingInL3} A, we present real space images of the emission (at a pumping power well above the lasing threshold) for a molecule (labeled M3) with inter-pillar distance $CC'=2.6\mu$m; in this molecule, lasing occurs in the $\ell=3$ states. We see that for a $\sigma_{+}$-polarized pump, the emission is strongly $\sigma_{+}$-polarized and presents six bright lobes localized at the center of each pillar with vanishing intensiy in-between. Interference patterns under $\sigma_{+}$ and $\sigma_{-}$ excitations are presented respectively in Fig. \ref{LasingInL3} B and D; the associated phase maps are reported in Fig. \ref{LasingInL3} C and E. The $\pi$ phase jump between adjacent pillars demonstrate that in molecule M3 lasing is triggered in the $\ell=3$ mode. As mentioned in the main text, this mode does not exhibit a net chirality, as it presents a mirror symmetry.
	
	\clearpage

	\section{Fine structure in n-pillar ring molecules}
	
	The discrete rotational symmetry of benzene photonic molecules allows defining a quantum number $\ell=0,\pm 1, \pm 2 ,3$ associated to the orbital angular momentum of eigenmodes which is related to their relative phase between the pillars. As described in the main text, the spin-orbit coupling of photons in these molecules, arising from the polarization-dependent hopping, splits the $\ell=\pm 1$ and $\ell=\pm 2$ manifolds to form 3-levels fine structures (see Fig. 1 (c) and (d) of the main text) that are necessary to implement our lasing scheme. Importantly, such a fine structure does not only appear in benzene molecules, but in any $n$-pillars ring molecules, with $n$ even and $>4$, allowing to extend the scheme to generate arbitrarily large OAM using molecules with a larger number of pillars.
	
	The point of this section is (1) to show how the emergence of this interesting fine structure can be understood in benzene molecules by considering the symmetry of the Hamiltonian, and (2) to extend this argument to molecules with arbitrary $n$. We demonstrate that they present the necessary 3-level fine structures in the $|\ell|=1$ and $|\ell|=n/2-1$ manifolds.
	
	\subsection{Fine structure in benzene molecules}
	
	The group of all symmetry operations of a benzene photonic molecule is $C_{6v}$. Contrary to real benzene molecules that present $D_{6}$ symmetry, our devices do not exhibit an out-of-plane mirror symmetry; this two groups are however isomorphic and lead to identical conclusions. The character table of $C_{6v}$ is presented in Table \ref{charC6v}, where $A_{i}$ and $B_{i}$ are unidimensional irreducible representations (irreps) of the group. $E_{i}$ are two-dimensional irreps; $E$ (identity), $C_{n}$ ($2\pi/n$ rotations) and $\sigma_{v}$ ($\sigma_{d}$) (reflections across vertical planes that cross two opposite pillars (links)) are the symmetry operations of the group.
	
	\begin{table}[h]
		\begin{tabular*}{0.5\textwidth}{@{\extracolsep{\fill} } l|c|c|cccccc}
			\cline{1-9}
			&&&&&&&&
			\\[-5pt]
			Modes & Functions & $C_{6v}$ & $E$ & $C_{6}$ & $C_{3}$ & $C_{2}$ & $\sigma_{v}$ & $\sigma_{d}$\\[2pt]
			\cline{1-9}
			&&&&&&&&
			\\[- 5 pt]
			$\ell=0$ & $z$ & $A_{1}$ & 1 & 1 & 1 & 1 & 1 & 1\\
			&& $A_{2}$ & 1 & 1 & 1 & 1 & -1 & -1\\
			$\ell=3$ & & $B_{1}$ & 1 & -1 & 1 & -1 & 1 & -1\\
			&& $B_{2}$ & 1 & -1 & 1 & -1 & -1 & 1\\
			$|\ell|=1$ & $(x,y)$ & $E_{1}$ & 2 & -2 & -1 & 1 & 0 & 0\\
			$|\ell|=2$ & & $E_{2}$ & 2 & 2 & -1 & -1 & 0 & 0
			\\[2 pt]
			\hline
			\multicolumn{2}{c}{} &&&&&&
			\\[-5 pt]
			\multicolumn{2}{c}{} & $\Gamma^{(\mathrm{scalar})}$ & 6 & 0 & 0 & 0 & 2 & 0
			\\[2 pt]
		\end{tabular*}
		\caption{Character table of the point group $C_{6v}$. The last line presents the characters of the reducible representation associated to scalar wave-functions of the benzene photonic molecule. The first and second columns respectively present scalar modes of the benzene and coordinates that transform according to the corresponding irreps presented in the third column.}
		\label{charC6v}
	\end{table}
	
	One possible basis for the scalar wave-functions of the molecule (i.e. the eigenmodes without considering the spin) is given by the ground state of each pillar ($\psi_{1-6}$); the eigenmodes of the systems can thus be expressed in the form of a vector $\Psi=(\psi_{1},\psi_{2},\psi_{3},\psi_{4},\psi_{5},\psi_{6})^{\dagger}$. One can then identify a set of matrices that describe how this vector transforms under each symmetry operation of the group.
	
	Since the Hamiltonian of the system commutes with every symmetry operator, this set of matrices forms a reducible representation of the group (that we label $\Gamma^{(\mathrm{scalar})}$). The trace (character) of each of these matrices is given in the bottom line of Table \ref{charC6v}, allowing to deduce the following decomposition in irreps:
	
	\begin{equation}
		\Gamma^{(\mathrm{scalar})}=A_{1} \oplus B_{1} \oplus E_{1} \oplus E_{2}.
	\end{equation}
	
	This decomposition indicates that the eigenspectrum of the scalar system is formed of 4 energy levels: 2 non-degenerate states transforming according to $A_{1}$ and $B_{1}$ and 2 degenerate doublets transforming as $E_{1}$ and $E_{2}$. 
	
\begin{figure*}[h]
		\includegraphics[trim=0cm 0cm 0cm 0cm, width=0.8\textwidth]{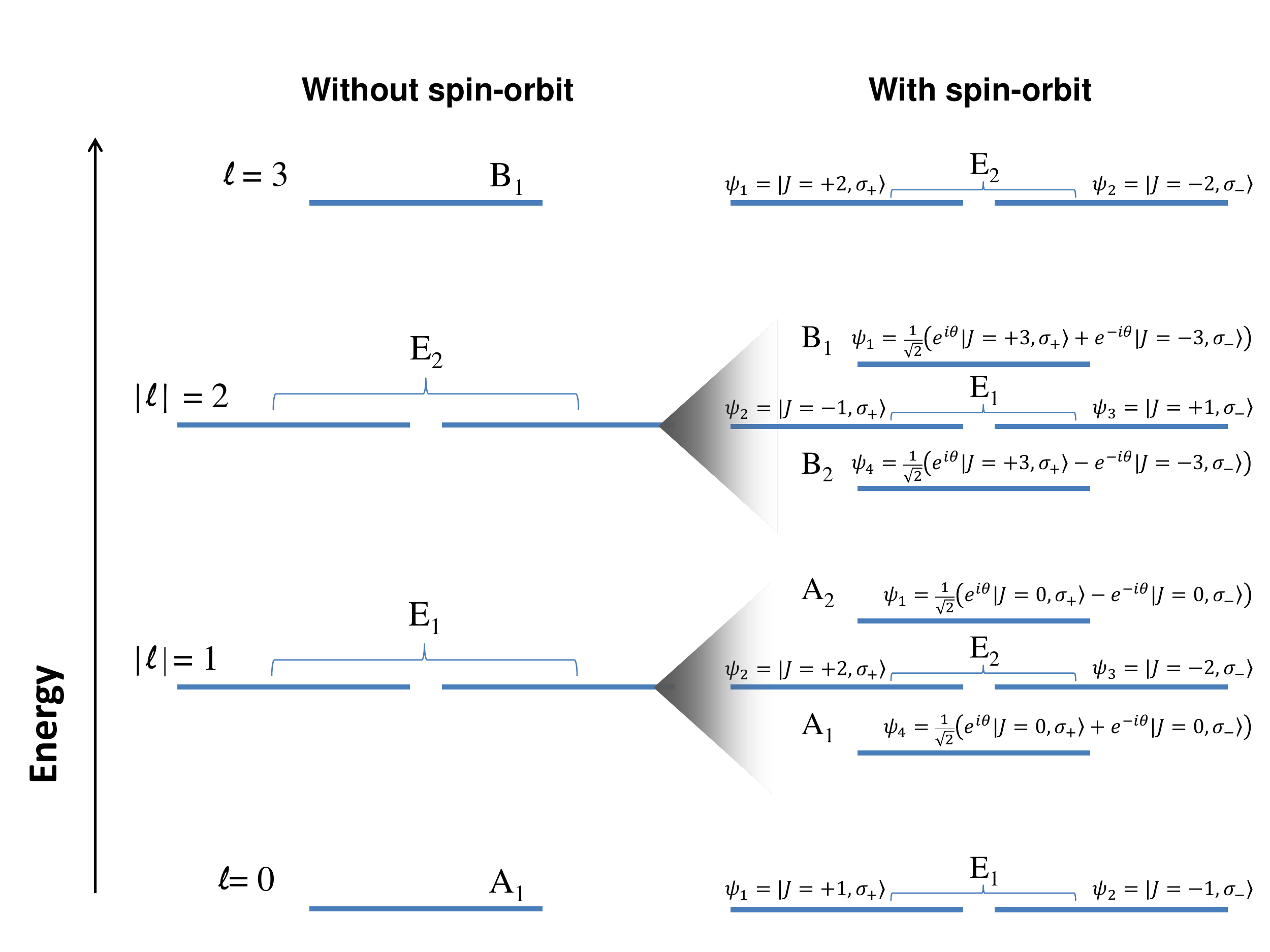}
		\caption{Energy levels fine structure of a benzene photonic molecule without (left) and with (right) spin-orbit coupling. The irreps and wave-function (in the basis $\ket{J,S}$) of each state is presented. Here, the phase $\theta=2\pi/6$.}
		\label{benzeneFS}
	\end{figure*}

	Inspection of the character table shows that the eigenmodes transforming according to $A_{1}$ and $B_{1}$ are, respectively, symmetric under all symmetry operations of the group (i.e. $\ell=0$) and anti-symmetric under $2\pi/6$ rotations (i.e. $\ell=3$). Furthermore, the doublets transforming as $E_{1}$ and $E_{2}$ correspond, respectively, to the $|\ell|= 1$ and $|\ell|= 2$ manifold (the value of the angular momentum is extracted directly from the wave-function of each mode, which are determined by expanding generators associated to each irreps, see Ref. [\onlinecite{Dresselhaus2008}]). Although group theory does not provide information on the relative energy of these different states, it is possible to spectrally order them, because their energy scales with the phase gradient associated to the angular momentum ($A_{1}-E_{1}-E_{2}-B_{1}$), as presented in the left part of Fig. \ref{benzeneFS}.
	
	In order to take into account the spin of photons, one needs to identify the irreps that transform identically as this spin moment: for the $C_{nv}$ symmetry groups, $\ell=\pm1$ angular momenta transform as $E_{1}$. Then, to retrieve the energy levels of our benzene molecule in the presence of spin-orbit coupling, we consider the tensor product between the $E_{1}$ and the scalar representation:
	
	\begin{align*}
		\label{decomposition}
		\Gamma^{(\mathrm{spin})} \otimes \Gamma^{(\mathrm{scalar})} &= E_{1}\otimes(A_{1} \oplus B_{1} \oplus E_{1} \oplus E_{2})\\
		&= \underbrace{E_{1}}_{\ell=0} \oplus \underbrace{E_{2}} _{\ell=3} \oplus \underbrace{A_{1} \oplus A_{2} \oplus E_{2}} _{|\ell|=1} \oplus \underbrace{B_{1} \oplus B_{2} \oplus E_{1}} _{|\ell|=2}.\addtocounter{equation}{1}\tag{\theequation}
	\end{align*}
	
	This decomposition shows the effect of spin-orbit coupling on the fine structure, which is summarized in Fig. \ref{benzeneFS}. In the spin-coupled basis, the $\ell=0$ and $\ell=3$ states become two-fold degenerate, due to the spin degeneracy, and transform respectively as $E_{1}$ and $E_{2}$ irreps. More importantly, the $\ell=1$ ($\ell=2$) manifolds split to form a 3-level fine structure formed from two non-degenerate states transforming as $A_{1}$ and $A_{2}$ for $|\ell|=1$ ($B_{1}$ and $B_{2}$ for $|\ell|=2$), and a doublet transforming as $E_{1}$ ($E_{2}$).
	
	By inspection of the symmetry of every eigenstate of the fine structure, it is possible to retrieve explicitly the wave-functions of each of these states. For the $\ell=1$ manifold, states that transform as $A_{1,2}$ are symmetric under $2\pi/6$ rotations, indicating that they carry a total angular momentum $\vec{J}=\vec{L}+\vec{S}=0$ (when considering spin-orbit coupling, $\ell$ and $S$ are no longer good quantum numbers, but $J$ is); furthermore, $A_{1}$ ($A_{2}$) is symmetric (anti-symmetric) under reflections $\sigma_{v,d}$. Therefore, the associated wave-functions (in the $\ket{J,S}$ basis) are:
	
	\begin{align*}
		\label{a1a2}
		\psi(A_{1})=\frac{1}{\sqrt{2}}[e^{i\theta}\ket{J=0,\sigma_{+}(\ell=-1)}+e^{-i\theta}\ket{J=0,\sigma_{-}(\ell=+1)}]\\
		\psi(A_{2})=\frac{1}{\sqrt{2}}[e^{i\theta}\ket{J=0,\sigma_{+}(\ell=-1)}-e^{-i\theta}\ket{J=0,\sigma_{-}(\ell=+1)}].\addtocounter{equation}{1}\tag{\theequation}
	\end{align*}
	
	\noindent in these equations, we have indicated the value of $\ell$ in each ket. Here, the phase $\theta=2\pi/6$ accounts for the 6-fold rotational symmetry of the wave-function.
	
	Wave-functions associated to the $E_{2}$ doublet form conjugated partners with a total angular momentum $J=2$, but opposite chiralities ($E_{2}$ is associated to modes presenting two quanta of angular momentum, see character table):
	
	\begin{equation}
		\psi_{\pm}(E_{2})=\ket{J=\pm 2,\sigma_{\pm}(\ell=\pm 1)}.
	\end{equation}
	
	In the $|\ell|=2$ manifold, states that transform according to $B_{1,2}$ are anti-symmetric under $2\pi/6$ rotations, indicating modes with total angular momentum $J=3$; furthermore, $B_{1}$ is symmetric (anti-symmetric) under $\sigma_{v}$ ($\sigma_{d}$) reflections (and vice-versa for $B_{2}$). Therefore, the associated wave-functions are:
	
	\begin{align*}
		\label{b1b2}
		\psi(B_{1})=\frac{1}{\sqrt{2}}[e^{i\theta}\ket{J=3,\sigma_{+}(\ell=+2)}+e^{-i\theta}\ket{J=-3,\sigma_{-}(\ell=-2)}]\\
		\psi(B_{2})=\frac{1}{\sqrt{2}}[e^{i\theta}\ket{J=3,\sigma_{+}(\ell=+2)}-e^{-i\theta}\ket{J=-3,\sigma_{-}(\ell=-2)}].\addtocounter{equation}{1}\tag{\theequation}
	\end{align*}
	
	Wave-functions associated to the $E_{1}$ doublet form conjugated partners with a total angular momentum $J=1$ ($E_{1}$ is the irrep associated to modes presenting one quantum of angular momentum):
	
	\begin{equation}
		\psi_{\pm}(E_{1})=\ket{J=\pm 1,\sigma_{\mp}(\ell=\pm 2)}.
	\end{equation}
	
	Importantly, each 3-level fine structure presents the followings necessary assets for our lasing scheme: (1) only one state is entirely polarized $\sigma_{+}$ and only one is entirely polarized $\sigma_{-}$, so that these states present the highest gain when spin-polarizing the gain medium, and (2) these $\sigma_{\pm}$-polarized states present opposite OAM (i.e. $\pm\ell$). This is indeed the case for the states forming the $E_{1}$ and $E_{2}$ doublets. Although group theory does not allow identifying the spectral ordering of the states forming both fine structures, this ordering does not impact whatsoever the implementation of the lasing scheme.
	
	Furthermore, finite elements numerical simulations have shown that symmetric and antisymmetric modes (i.e. $A_{1}$ and $A_{2}$ for $|\ell|=1$, and $B_{1}$ and $B_{2}$ for $|\ell|=2$) are maximally separated in energy (the doublets $E_{1,2}$ falling in-between); these simulations as well show that the highest energy state has $A_{2}$ ($B_{1}$) symmetry for the $|\ell|=1$ ($|\ell|=2$) manifold. The energy splitting, in both cases, is proportional to the hopping energy difference for photons polarized along and perpendicularly to the axis linking neighbouring pillars.
	
	\subsection{Extension to arbitrary n-pillar molecules}
	
	We now consider the general case of a n-pillar molecule whose symmetry elements form an ensemble corresponding to the $C_{nv}$ group. Table \ref{charCnvEven} and  \ref{charCnvOdd} present the character table for arbitrary $n$, even and odd respectively.\\
	
	\textbf{Even number of pillars ($\mathrm{n>4}$)}
	
	\begin{table}[h]
		\begin{tabular*}{\textwidth}{@{\extracolsep{\fill} } l|c|c|cccccccc}
			\cline{1-11}
			&&&&&&&&&&
			\\[-5pt]
			\multirow{2}{*}{Modes} & \multirow{2}{*}{Functions} & $C_{nv}$ & \multirow{2}{*}{$E$} & \multirow{2}{*}{$2C_{n}$} & \multirow{2}{*}{$2C_{n}^{2}$} & \multirow{2}{*}{$2C_{n}^{3}$} & \multirow{2}{*}{...} & \multirow{2}{*}{$C_{n}^{n/2}$} & \multirow{2}{*}{$3\sigma_{v}$} & \multirow{2}{*}{$3\sigma_{d}$}\\
			&& ($n$ even)&&&&&&&&\\[2pt]
			\cline{1-11}
			&&&&&&&&&&
			\\[- 5 pt]
			$|\ell|=0$ & $z$ & $A_{1}$ & 1 & 1 & 1 & 1 & ... & 1 & 1 & 1 \\
			&& $A_{2}$ & 1 & 1 & 1 & 1 & ... & 1 & -1 & -1 \\
			$|\ell|=n/2$ & & $B_{1}$ & 1 & -1 & 1 & -1 & ... & -1 & 1 & -1 \\
			&& $B_{2}$ & 1 & -1 & 1 & -1 & ... & -1 & -1 & 1 \\
			$|\ell|=1$ & $(x,y)$ & $E_{1}$ & 2 & 2cos($\alpha$) & 2cos(2$\alpha$) & 2cos(3$\alpha$) & ... & 2cos($\tfrac{n}{2}\alpha$) & 0 & 0 \\
			$|\ell|=2$ &  & $E_{2}$ & 2 & 2cos($2\alpha$) & 2cos(4$\alpha$) & 2cos(6$\alpha$) & ... & 2cos($n\alpha$) & 0 & 0 \\
			$\ell=3$ &  & $E_{3}$ & 2 & 2cos($3\alpha$) & 2cos(6$\alpha$) & 2cos(9$\alpha$) & ... & 2cos($\tfrac{3n}{2}\alpha$) & 0 & 0 \\
			...&...&...&...&...&...&...&...&...&...&...\\
			$|\ell|=n/2-1$ &  & $E_{n/2-1}$ & 2 & 2cos($(\tfrac{n}{2}-1)\alpha$) & 2cos($(n-2)\alpha$) &  2cos($(\tfrac{3n}{2}-3)\alpha$) & ... &  2cos($\tfrac{n}{2}(\tfrac{n}{2}-1)\alpha$) & 0 & 0 
			\\[2 pt]
			\hline
			\multicolumn{2}{c}{} &&&&&&&&
			\\[-5 pt]
			\multicolumn{2}{c}{} & $\Gamma^{(\mathrm{scalar})}$ & n & 0 & 0 & 0 & ... & 0 & 2 & 0
			\\[2 pt]
		\end{tabular*}
		\caption{General character table of a point group $C_{nv}$ with n even. The bottom line corresponds to the characters of the reducible representation associated to the scalar wave-functions (i.e. without spin). ($\alpha=2\pi/n$)}
		\label{charCnvEven}
	\end{table}
	
	For a molecule with an even number of pillars (but $n>4$), the decomposition of the representation associated to the scalar wave-functions (i.e. without spin) is:
	
	\begin{equation}\label{decompEven}
		\Gamma^{(\mathrm{scalar})}_{\mathrm{even}}=A_{1} \oplus B_{1} \oplus E_{1} \oplus E_{2} \oplus E_{3} \oplus ... \oplus E_{n/2-1}.
	\end{equation}
	
\begin{figure*}[h]
		\includegraphics[trim=0cm 0cm 0cm 0cm, width=0.8\textwidth]{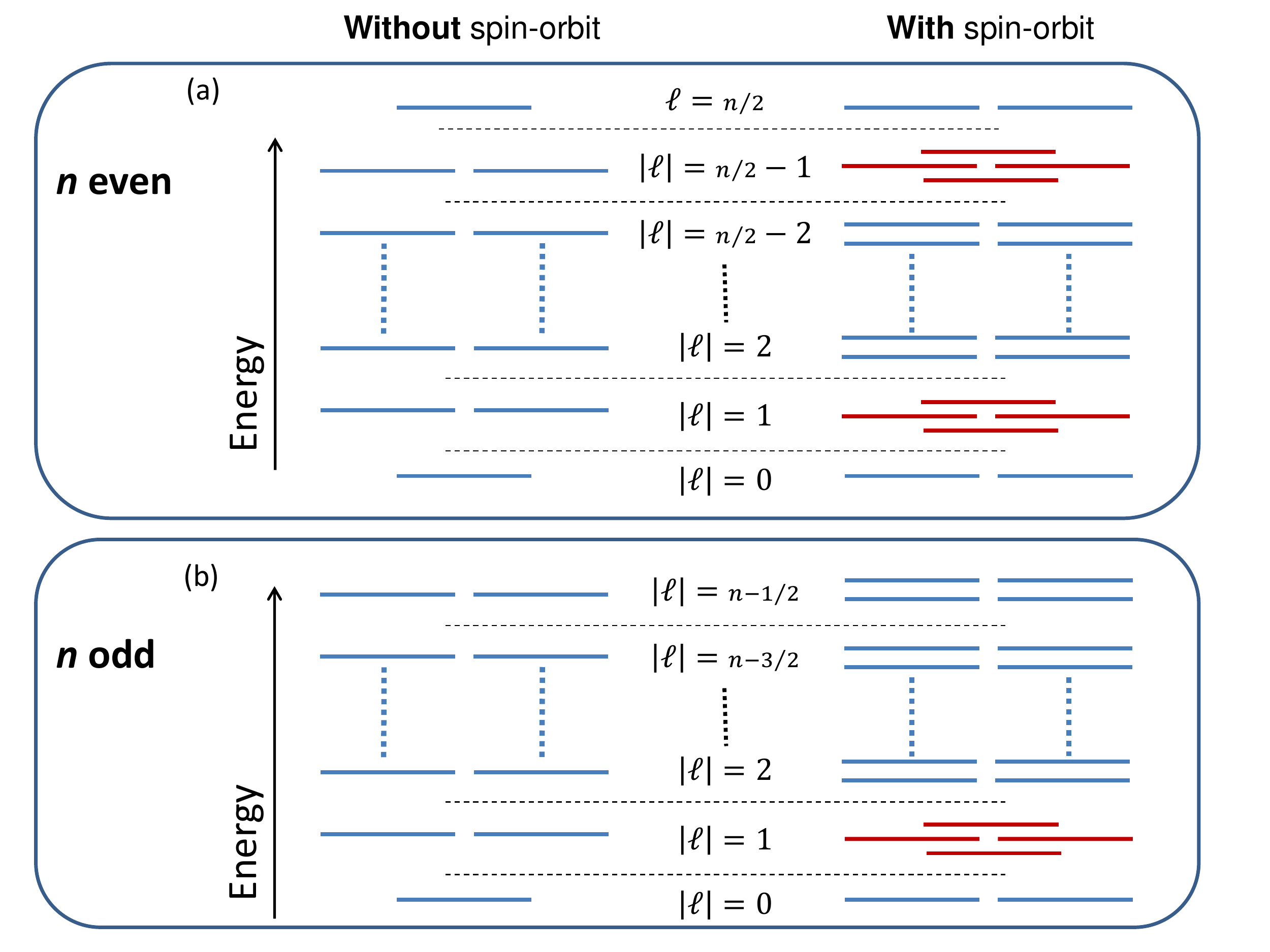}
		\caption{Energy levels fine structure of a n-pillar photonic molecule for n even (top) and odd (bottom), with and without. The 3-level fine structure necessary to implement our lasing scheme are presented in red.}
		\label{even_odd_FS}
	\end{figure*}

	This indicates that the energy level structure is formed from 2 non-degenerate modes: $A_{1}$ ($\ell=0$) and $B_{1}$ ($\ell=n/2$); and $n/2-1$ doublets associated to increasing $|\ell|$. A schematic representation of this level structure is presented in the left part of Fig. \ref{even_odd_FS} (a).
	
	To take into account the effect of spin-orbit coupling, we take the tensor product between the representations associated to the scalar wave-functions and $E_{1}$, corresponding to the spin moment. The decomposition of this tensor product gives:
	
	\begin{align*}
		E_{1} \otimes \Gamma^{(\mathrm{scalar})}_{\mathrm{even}} &= E_{1} \otimes (A_{1} \oplus B_{1} \oplus E_{1} \oplus E_{2} \oplus E_{3} \oplus ... \oplus E_{n/2-1})\\
		&= \underbrace{E_{1}} _{\ell=0} \oplus \underbrace{E_{n/2-1}} _{\ell=3} \oplus \underbrace{A_{1} \oplus A_{2} \oplus E_{2}} _{|\ell|=1} \oplus \underbrace{E_{1} \oplus E_{3}} _{|\ell|=2} \oplus  \underbrace{E_{2} \oplus E_{4}} _{\ell=3} \oplus ... \oplus \underbrace{E_{n/2-3} \oplus E_{n/2-1}} _{|\ell|=4} \oplus \underbrace{B_{1} \oplus B_{2} \oplus E_{n/2-2}} _{|\ell|=n/2-1}. 
		\addtocounter{equation}{1}\tag{\theequation}
	\end{align*}
	
	This decomposition (schematically presented in Fig. \ref{even_odd_FS} (a)) shows that the $\ell=0$ and $\ell=n/2$ manifolds, as in benzene, form degenerate doublets transforming as $E_{1}$ and $E_{n/2-1}$, respectively. Importantly, the $|\ell|=1$ and $|\ell|=n/2-1$ manifolds split respectively in 3-level fine structures similar to the $|\ell|=1$ and $|\ell=2$ manifolds of benzene. Therefore, both manifolds (indicated in red in Fig. \ref{even_odd_FS} (a)) offer an appropriate fine structure for implementing our OAM lasing scheme.
	
	The other manifolds of the eigenspectrum ($2\le |\ell| \le n/2-2$) split in pairs of doublets transforming as $E_{|\ell|-1}$ and $E_{|\ell|+1}$. The associated wave-functions (with corresponding OAM) can be written as:
	
	\begin{align*}
		\psi_{\pm}(E_{|\ell|-1})&=\ket{J=\pm(|\ell|+1), \sigma_{\mp}}\rightarrow \mathrm{OAM}=\pm|\ell|\\
		\psi_{\pm}(E_{|\ell|+1})&=\ket{J=\pm(|\ell|-1), \sigma_{\pm}}\rightarrow \mathrm{OAM}=\pm|\ell|.  \addtocounter{equation}{1}\tag{\theequation}
	\end{align*}
	
	Consequently, none of these manifolds offer an appropriate fine structure to implement our lasing scheme. Indeed, although all the states are circularly polarized, the polarization ($\sigma_{\pm}$) is not linked to a single chirality, thus preventing triggering lasing in a mode carrying a net OAM when spin-polarizing the gain medium.\\
	
	\textbf{4-pillar molecules}
	
	For a 4-pillar molecule, the decomposition described in Eq. \eqref{decompEven} breaks down. In such a structure, $\ell=\pm1$ modes transform according to the $E_{1}$ irrep of $C_{2v}$. Taking into account spin-orbit coupling, these states splits according to the decomposition:
	
	\begin{equation}
		E_{1}\otimes E_{1}=A_{1}\oplus A_{2} \oplus B_{1} \oplus B_{2},
	\end{equation}
	
	\noindent thus forming 4 linearly polarized states carrying no net OAM. A possible way to understand this result in our particular system is that the symmetry of the square architecture does not lead to a mixing of the linear polarization when photons jump from one pillar to another.\\
	
	\textbf{Odd number of pillars}
	
	\begin{table}[h]
		\begin{tabular*}{\textwidth}{@{\extracolsep{\fill} } l|c|c|cccccccc}
			\cline{1-11}
			&&&&&&&&&&
			\\[-5pt]
			\multirow{2}{*}{Modes} & \multirow{2}{*}{Functions} & $C_{nv}$ & \multirow{2}{*}{$E$} & \multirow{2}{*}{$2C_{n}$} & \multirow{2}{*}{$2C_{n}^{2}$} & \multirow{2}{*}{$2C_{n}^{3}$} & \multirow{2}{*}{...} & \multirow{2}{*}{$2C_{n}^{\tfrac{n-1}{2}}$} & \multirow{2}{*}{$3\sigma_{v}$} & \multirow{2}{*}{$3\sigma_{d}$}\\
			&& ($n$ odd)&&&&&&&&\\[2pt]
			\cline{1-11}
			&&&&&&&&&&
			\\[- 5 pt]
			$|\ell|=0$ & $z$ & $A_{1}$ & 1 & 1 & 1 & 1 & ... & 1 & 1 & 1 \\
			&& $A_{2}$ & 1 & 1 & 1 & 1 & ... & 1 & -1 & -1 \\
			$|\ell|=1$ & $(x,y)$ & $E_{1}$ & 2 & 2cos($\alpha$) & 2cos(2$\alpha$) & 2cos(3$\alpha$) & ... & 2cos($\tfrac{n-1}{2}\alpha$) & 0 & 0 \\
			$|\ell|=2$ && $E_{2}$ & 2 & 2cos($2\alpha$) & 2cos(4$\alpha$) & 2cos(6$\alpha$) & ... & 2cos($(n-1)\alpha$) & 0 & 0 \\
			$\ell=3$ && $E_{3}$ & 2 & 2cos($3\alpha$) & 2cos(6$\alpha$) & 2cos(9$\alpha$) & ... & 2cos($\tfrac{3(n-1)}{2}\alpha$) & 0 & 0 \\
			...&...&...&...&...&...&...&...&...&...&...\\
			$|\ell|=(n-1)/2$ &  & $E_{(n-1)/2}$ & 2 & 2cos($(\tfrac{n-1}{2})\alpha$) & 2cos($(n-1)\alpha$) &  2cos($(\tfrac{3(n-1)}{2})\alpha$) & ... &  2cos($(\tfrac{n-1}{2})^{2}\alpha$) & 0 & 0 
			\\[2 pt]
			\hline
			\multicolumn{2}{c}{} &&&&&&&&
			\\[-5 pt]
			\multicolumn{2}{c}{} & $\Gamma^{(\mathrm{scalar})}$ & n & 0 & 0 & 0 & ... & 0 & 2 & 0
			\\[2 pt]
		\end{tabular*}
		\caption{General character table of a point group $C_{nv}$ with n odd. The bottom line corresponds to the characters of the reducible representation associated to the scalar wave-functions (i.e. without spin). ($\alpha=2\pi/n$)}
		\label{charCnvOdd}
	\end{table}
	
	For a molecule presenting an odd number of pillars, the representation associated to scalar wave-functions is decomposed in irreps as:
	
	\begin{equation}
		\Gamma^{(\mathrm{scalar})}_{\mathrm{odd}}=A_{1}\oplus E_{1} \oplus E_{2} \oplus E_{3} \oplus ... \oplus E_{(n-1)/2}.
	\end{equation}
	
	The main difference with the case of even molecules is the disappearance of the $B_{1}$ states, as an antisymmetric state (i.e. presenting a $\pi$ phase shift between neighbouring pillars) is impossible in an odd number of pillars. A schematic representation of this level structure is presented in the left part of Fig. \ref{even_odd_FS} (b).
	
	When including the spin degree of freedom, the eigenspectrum (schematically depicted in the right part of Fig. \ref{even_odd_FS} (b))can be decomposed as:
	
	\begin{align*}
		E_{1} \otimes \Gamma^{(\mathrm{scalar})}_{\mathrm{odd}} &= E_{1} \otimes (A_{1} \oplus E_{1} \oplus E_{2} \oplus E_{3} \oplus ... \oplus E_{(n-1)/2})\\
		&= \underbrace{E_{1}} _{\ell=0} \oplus \underbrace{A_{1} \oplus A_{2} \oplus E_{2}} _{|\ell|=1} \oplus \underbrace{E_{1} \oplus E_{3}} _{|\ell|=2} \oplus  \underbrace{E_{2} \oplus E_{4}} _{\ell=3} \oplus ... \oplus \underbrace{E_{(n-3)/2} \oplus E_{(n-1)/2}} _{|\ell|=(n-1)/2}. \tag{\theequation}
	\end{align*}
	
	Here, contrary to the case of an even number of pillars, only the $|\ell|=1$ manifold exhibits the appropriate 3-level structure (indicated in red in Fig. \ref{even_odd_FS} (b)); all the other manifolds split in pairs of doublets that does not allow implementing our lasing scheme.
	
\begin{thebibliography}{41}%
		\makeatletter
		\providecommand \@ifxundefined [1]{%
		\@ifx{#1\undefined}
		}%
		\providecommand \@ifnum [1]{%
		\ifnum #1\expandafter \@firstoftwo
		\else \expandafter \@secondoftwo
		\fi
		}%
		\providecommand \@ifx [1]{%
		\ifx #1\expandafter \@firstoftwo
		\else \expandafter \@secondoftwo
		\fi
		}%
		\providecommand \natexlab [1]{#1}%
		\providecommand \enquote  [1]{``#1''}%
		\providecommand \bibnamefont  [1]{#1}%
		\providecommand \bibfnamefont [1]{#1}%
		\providecommand \citenamefont [1]{#1}%
		\providecommand \href@noop [0]{\@secondoftwo}%
		\providecommand \href [0]{\begingroup \@sanitize@url \@href}%
		\providecommand \@href[1]{\@@startlink{#1}\@@href}%
		\providecommand \@@href[1]{\endgroup#1\@@endlink}%
		\providecommand \@sanitize@url [0]{\catcode `\\12\catcode `\$12\catcode
		`\&12\catcode `\#12\catcode `\^12\catcode `\_12\catcode `\%12\relax}%
		\providecommand \@@startlink[1]{}%
		\providecommand \@@endlink[0]{}%
		\providecommand \url  [0]{\begingroup\@sanitize@url \@url }%
		\providecommand \@url [1]{\endgroup\@href {#1}{\urlprefix }}%
		\providecommand \urlprefix  [0]{URL }%
		\providecommand \Eprint [0]{\href }%
		\providecommand \doibase [0]{http://dx.doi.org/}%
		\providecommand \selectlanguage [0]{\@gobble}%
		\providecommand \bibinfo  [0]{\@secondoftwo}%
		\providecommand \bibfield  [0]{\@secondoftwo}%
		\providecommand \translation [1]{[#1]}%
		\providecommand \BibitemOpen [0]{}%
		\providecommand \bibitemStop [0]{}%
		\providecommand \bibitemNoStop [0]{.\EOS\space}%
		\providecommand \EOS [0]{\spacefactor3000\relax}%
		\providecommand \BibitemShut  [1]{\csname bibitem#1\endcsname}%
		\let\auto@bib@innerbib\@empty
		\bibitem [{\citenamefont {Allen}\ \emph {et~al.}(1992)\citenamefont {Allen},
		\citenamefont {Beijersbergen}, \citenamefont {Spreeuw},\ and\ \citenamefont
		{Woerdman}}]{Allen1992}%
		\BibitemOpen
		\bibfield  {author} {\bibinfo {author} {\bibfnamefont {L.}~\bibnamefont
		{Allen}}, \bibinfo {author} {\bibfnamefont {M.~W.}\ \bibnamefont
		{Beijersbergen}}, \bibinfo {author} {\bibfnamefont {R.~J.~C.}\ \bibnamefont
		{Spreeuw}}, \ and\ \bibinfo {author} {\bibfnamefont {J.~P.}\ \bibnamefont
		{Woerdman}},\ }\href {\doibase 10.1103/PhysRevA.45.8185} {\bibfield
		{journal} {\bibinfo  {journal} {Physical Review A}\ }\textbf {\bibinfo
		{volume} {45}},\ \bibinfo {pages} {8185} (\bibinfo {year}
		{1992})}\BibitemShut {NoStop}%
		\bibitem [{\citenamefont {Fickler}\ \emph {et~al.}(2016)\citenamefont
		{Fickler}, \citenamefont {Campbell}, \citenamefont {Buchler}, \citenamefont
		{Lam},\ and\ \citenamefont {Zeilinger}}]{Fickler2016}%
		\BibitemOpen
		\bibfield  {author} {\bibinfo {author} {\bibfnamefont {R.}~\bibnamefont
		{Fickler}}, \bibinfo {author} {\bibfnamefont {G.}~\bibnamefont {Campbell}},
		\bibinfo {author} {\bibfnamefont {B.}~\bibnamefont {Buchler}}, \bibinfo
		{author} {\bibfnamefont {P.~K.}\ \bibnamefont {Lam}}, \ and\ \bibinfo
		{author} {\bibfnamefont {A.}~\bibnamefont {Zeilinger}},\ }\href {\doibase
		10.1073/pnas.1616889113} {\bibfield  {journal} {\bibinfo  {journal}
		{Proceedings of the National Academy of Sciences of the United States of
		America}\ }\textbf {\bibinfo {volume} {113}},\ \bibinfo {pages} {13642}
		(\bibinfo {year} {2016})}\BibitemShut {NoStop}%
		\bibitem [{\citenamefont {Gibson}\ \emph {et~al.}(2004)\citenamefont {Gibson},
		\citenamefont {Courtial}, \citenamefont {Padgett}, \citenamefont {Vasnetsov},
		\citenamefont {Pas'ko}, \citenamefont {Barnett},\ and\ \citenamefont
		{Franke-Arnold}}]{Gibson2004}%
		\BibitemOpen
		\bibfield  {author} {\bibinfo {author} {\bibfnamefont {G.}~\bibnamefont
		{Gibson}}, \bibinfo {author} {\bibfnamefont {J.}~\bibnamefont {Courtial}},
		\bibinfo {author} {\bibfnamefont {M.~J.}\ \bibnamefont {Padgett}}, \bibinfo
		{author} {\bibfnamefont {M.}~\bibnamefont {Vasnetsov}}, \bibinfo {author}
		{\bibfnamefont {V.}~\bibnamefont {Pas'ko}}, \bibinfo {author} {\bibfnamefont
		{S.~M.}\ \bibnamefont {Barnett}}, \ and\ \bibinfo {author} {\bibfnamefont
		{S.}~\bibnamefont {Franke-Arnold}},\ }\href {\doibase 10.1364/OPEX.12.005448}
		{\bibfield  {journal} {\bibinfo  {journal} {Optics Express}\ }\textbf
		{\bibinfo {volume} {12}},\ \bibinfo {pages} {5448} (\bibinfo {year}
		{2004})}\BibitemShut {NoStop}%
		\bibitem [{\citenamefont {Wang}\ \emph {et~al.}(2012)\citenamefont {Wang},
		\citenamefont {Yang}, \citenamefont {Fazal}, \citenamefont {Ahmed},
		\citenamefont {Yan}, \citenamefont {Huang}, \citenamefont {Ren},
		\citenamefont {Yue}, \citenamefont {Dolinar}, \citenamefont {Tur},\ and\
		\citenamefont {Willner}}]{Wang2012}%
		\BibitemOpen
		\bibfield  {author} {\bibinfo {author} {\bibfnamefont {J.}~\bibnamefont
		{Wang}}, \bibinfo {author} {\bibfnamefont {J.-Y.}\ \bibnamefont {Yang}},
		\bibinfo {author} {\bibfnamefont {I.~M.}\ \bibnamefont {Fazal}}, \bibinfo
		{author} {\bibfnamefont {N.}~\bibnamefont {Ahmed}}, \bibinfo {author}
		{\bibfnamefont {Y.}~\bibnamefont {Yan}}, \bibinfo {author} {\bibfnamefont
		{H.}~\bibnamefont {Huang}}, \bibinfo {author} {\bibfnamefont
		{Y.}~\bibnamefont {Ren}}, \bibinfo {author} {\bibfnamefont {Y.}~\bibnamefont
		{Yue}}, \bibinfo {author} {\bibfnamefont {S.}~\bibnamefont {Dolinar}},
		\bibinfo {author} {\bibfnamefont {M.}~\bibnamefont {Tur}}, \ and\ \bibinfo
		{author} {\bibfnamefont {A.~E.}\ \bibnamefont {Willner}},\ }\href {\doibase
		10.1038/nphoton.2012.138} {\bibfield  {journal} {\bibinfo  {journal} {Nature
		Photonics}\ }\textbf {\bibinfo {volume} {6}},\ \bibinfo {pages} {488}
		(\bibinfo {year} {2012})}\BibitemShut {NoStop}%
		\bibitem [{\citenamefont {Bozinovic}\ \emph {et~al.}(2013)\citenamefont
		{Bozinovic}, \citenamefont {Yue}, \citenamefont {Ren}, \citenamefont {Tur},
		\citenamefont {Kristensen}, \citenamefont {Huang}, \citenamefont {Willner},\
		and\ \citenamefont {Ramachandran}}]{Bozinovic2013}%
		\BibitemOpen
		\bibfield  {author} {\bibinfo {author} {\bibfnamefont {N.}~\bibnamefont
		{Bozinovic}}, \bibinfo {author} {\bibfnamefont {Y.}~\bibnamefont {Yue}},
		\bibinfo {author} {\bibfnamefont {Y.}~\bibnamefont {Ren}}, \bibinfo {author}
		{\bibfnamefont {M.}~\bibnamefont {Tur}}, \bibinfo {author} {\bibfnamefont
		{P.}~\bibnamefont {Kristensen}}, \bibinfo {author} {\bibfnamefont
		{H.}~\bibnamefont {Huang}}, \bibinfo {author} {\bibfnamefont {A.~E.}\
		\bibnamefont {Willner}}, \ and\ \bibinfo {author} {\bibfnamefont
		{S.}~\bibnamefont {Ramachandran}},\ }\href {\doibase 10.1126/science.1237861}
		{\bibfield  {journal} {\bibinfo  {journal} {Science (New York, N.Y.)}\
		}\textbf {\bibinfo {volume} {340}},\ \bibinfo {pages} {1545} (\bibinfo {year}
		{2013})}\BibitemShut {NoStop}%
		\bibitem [{\citenamefont {Willner}\ \emph {et~al.}(2015)\citenamefont
		{Willner}, \citenamefont {Huang}, \citenamefont {Yan}, \citenamefont {Ren},
		\citenamefont {Ahmed}, \citenamefont {Xie}, \citenamefont {Bao},
		\citenamefont {Li}, \citenamefont {Cao}, \citenamefont {Zhao}, \citenamefont
		{Wang}, \citenamefont {Lavery}, \citenamefont {Tur}, \citenamefont
		{Ramachandran}, \citenamefont {Molisch}, \citenamefont {Ashrafi},\ and\
		\citenamefont {Ashrafi}}]{Willner2015}%
		\BibitemOpen
		\bibfield  {author} {\bibinfo {author} {\bibfnamefont {A.~E.}\ \bibnamefont
		{Willner}}, \bibinfo {author} {\bibfnamefont {H.}~\bibnamefont {Huang}},
		\bibinfo {author} {\bibfnamefont {Y.}~\bibnamefont {Yan}}, \bibinfo {author}
		{\bibfnamefont {Y.}~\bibnamefont {Ren}}, \bibinfo {author} {\bibfnamefont
		{N.}~\bibnamefont {Ahmed}}, \bibinfo {author} {\bibfnamefont
		{G.}~\bibnamefont {Xie}}, \bibinfo {author} {\bibfnamefont {C.}~\bibnamefont
		{Bao}}, \bibinfo {author} {\bibfnamefont {L.}~\bibnamefont {Li}}, \bibinfo
		{author} {\bibfnamefont {Y.}~\bibnamefont {Cao}}, \bibinfo {author}
		{\bibfnamefont {Z.}~\bibnamefont {Zhao}}, \bibinfo {author} {\bibfnamefont
		{J.}~\bibnamefont {Wang}}, \bibinfo {author} {\bibfnamefont {M.~P.~J.}\
		\bibnamefont {Lavery}}, \bibinfo {author} {\bibfnamefont {M.}~\bibnamefont
		{Tur}}, \bibinfo {author} {\bibfnamefont {S.}~\bibnamefont {Ramachandran}},
		\bibinfo {author} {\bibfnamefont {A.~F.}\ \bibnamefont {Molisch}}, \bibinfo
		{author} {\bibfnamefont {N.}~\bibnamefont {Ashrafi}}, \ and\ \bibinfo
		{author} {\bibfnamefont {S.}~\bibnamefont {Ashrafi}},\ }\href {\doibase
		10.1364/AOP.7.000066} {\bibfield  {journal} {\bibinfo  {journal} {Advances in
		Optics and Photonics}\ }\textbf {\bibinfo {volume} {7}},\ \bibinfo {pages}
		{66} (\bibinfo {year} {2015})}\BibitemShut {NoStop}%
		\bibitem [{\citenamefont {Gr{\"{o}}blacher}\ \emph {et~al.}(2006)\citenamefont
		{Gr{\"{o}}blacher}, \citenamefont {Jennewein}, \citenamefont {Vaziri},
		\citenamefont {Weihs},\ and\ \citenamefont {Zeilinger}}]{Groblacher2006}%
		\BibitemOpen
		\bibfield  {author} {\bibinfo {author} {\bibfnamefont {S.}~\bibnamefont
		{Gr{\"{o}}blacher}}, \bibinfo {author} {\bibfnamefont {T.}~\bibnamefont
		{Jennewein}}, \bibinfo {author} {\bibfnamefont {A.}~\bibnamefont {Vaziri}},
		\bibinfo {author} {\bibfnamefont {G.}~\bibnamefont {Weihs}}, \ and\ \bibinfo
		{author} {\bibfnamefont {A.}~\bibnamefont {Zeilinger}},\ }\href {\doibase
		10.1088/1367-2630/8/5/075} {\bibfield  {journal} {\bibinfo  {journal} {New
		Journal of Physics}\ }\textbf {\bibinfo {volume} {8}},\ \bibinfo {pages} {75}
		(\bibinfo {year} {2006})}\BibitemShut {NoStop}%
		\bibitem [{\citenamefont {Vallone}\ \emph {et~al.}(2014)\citenamefont
		{Vallone}, \citenamefont {D’Ambrosio}, \citenamefont {Sponselli},
		\citenamefont {Slussarenko}, \citenamefont {Marrucci}, \citenamefont
		{Sciarrino},\ and\ \citenamefont {Villoresi}}]{Vallone2014}%
		\BibitemOpen
		\bibfield  {author} {\bibinfo {author} {\bibfnamefont {G.}~\bibnamefont
		{Vallone}}, \bibinfo {author} {\bibfnamefont {V.}~\bibnamefont
		{D’Ambrosio}}, \bibinfo {author} {\bibfnamefont {A.}~\bibnamefont
		{Sponselli}}, \bibinfo {author} {\bibfnamefont {S.}~\bibnamefont
		{Slussarenko}}, \bibinfo {author} {\bibfnamefont {L.}~\bibnamefont
		{Marrucci}}, \bibinfo {author} {\bibfnamefont {F.}~\bibnamefont {Sciarrino}},
		\ and\ \bibinfo {author} {\bibfnamefont {P.}~\bibnamefont {Villoresi}},\
		}\href {\doibase 10.1103/PhysRevLett.113.060503} {\bibfield  {journal}
		{\bibinfo  {journal} {Physical Review Letters}\ }\textbf {\bibinfo {volume}
		{113}},\ \bibinfo {pages} {060503} (\bibinfo {year} {2014})}\BibitemShut
		{NoStop}%
		\bibitem [{\citenamefont {Sit}\ \emph {et~al.}(2017)\citenamefont {Sit},
		\citenamefont {Bouchard}, \citenamefont {Fickler}, \citenamefont
		{Gagnon-Bischoff}, \citenamefont {Larocque}, \citenamefont {Heshami},
		\citenamefont {Elser}, \citenamefont {Peuntinger}, \citenamefont
		{G{\"{u}}nthner}, \citenamefont {Heim}, \citenamefont {Marquardt},
		\citenamefont {Leuchs}, \citenamefont {Boyd},\ and\ \citenamefont
		{Karimi}}]{Sit2017}%
		\BibitemOpen
		\bibfield  {author} {\bibinfo {author} {\bibfnamefont {A.}~\bibnamefont
		{Sit}}, \bibinfo {author} {\bibfnamefont {F.}~\bibnamefont {Bouchard}},
		\bibinfo {author} {\bibfnamefont {R.}~\bibnamefont {Fickler}}, \bibinfo
		{author} {\bibfnamefont {J.}~\bibnamefont {Gagnon-Bischoff}}, \bibinfo
		{author} {\bibfnamefont {H.}~\bibnamefont {Larocque}}, \bibinfo {author}
		{\bibfnamefont {K.}~\bibnamefont {Heshami}}, \bibinfo {author} {\bibfnamefont
		{D.}~\bibnamefont {Elser}}, \bibinfo {author} {\bibfnamefont
		{C.}~\bibnamefont {Peuntinger}}, \bibinfo {author} {\bibfnamefont
		{K.}~\bibnamefont {G{\"{u}}nthner}}, \bibinfo {author} {\bibfnamefont
		{B.}~\bibnamefont {Heim}}, \bibinfo {author} {\bibfnamefont {C.}~\bibnamefont
		{Marquardt}}, \bibinfo {author} {\bibfnamefont {G.}~\bibnamefont {Leuchs}},
		\bibinfo {author} {\bibfnamefont {R.~W.}\ \bibnamefont {Boyd}}, \ and\
		\bibinfo {author} {\bibfnamefont {E.}~\bibnamefont {Karimi}},\ }\href
		{\doibase 10.1364/OPTICA.4.001006} {\bibfield  {journal} {\bibinfo  {journal}
		{Optica}\ }\textbf {\bibinfo {volume} {4}},\ \bibinfo {pages} {1006}
		(\bibinfo {year} {2017})}\BibitemShut {NoStop}%
		\bibitem [{\citenamefont {Wang}\ \emph
		{et~al.}(2018{\natexlab{a}})\citenamefont {Wang}, \citenamefont {Paesani},
		\citenamefont {Ding}, \citenamefont {Santagati}, \citenamefont {Skrzypczyk},
		\citenamefont {Salavrakos}, \citenamefont {Tura}, \citenamefont {Augusiak},
		\citenamefont {Man{\v{c}}inska}, \citenamefont {Bacco}, \citenamefont
		{Bonneau}, \citenamefont {Silverstone}, \citenamefont {Gong}, \citenamefont
		{Ac{\'{\i}}n}, \citenamefont {Rottwitt}, \citenamefont {Oxenl{\o}we},
		\citenamefont {O'Brien}, \citenamefont {Laing},\ and\ \citenamefont
		{Thompson}}]{Wang2018}%
		\BibitemOpen
		\bibfield  {author} {\bibinfo {author} {\bibfnamefont {J.}~\bibnamefont
		{Wang}}, \bibinfo {author} {\bibfnamefont {S.}~\bibnamefont {Paesani}},
		\bibinfo {author} {\bibfnamefont {Y.}~\bibnamefont {Ding}}, \bibinfo {author}
		{\bibfnamefont {R.}~\bibnamefont {Santagati}}, \bibinfo {author}
		{\bibfnamefont {P.}~\bibnamefont {Skrzypczyk}}, \bibinfo {author}
		{\bibfnamefont {A.}~\bibnamefont {Salavrakos}}, \bibinfo {author}
		{\bibfnamefont {J.}~\bibnamefont {Tura}}, \bibinfo {author} {\bibfnamefont
		{R.}~\bibnamefont {Augusiak}}, \bibinfo {author} {\bibfnamefont
		{L.}~\bibnamefont {Man{\v{c}}inska}}, \bibinfo {author} {\bibfnamefont
		{D.}~\bibnamefont {Bacco}}, \bibinfo {author} {\bibfnamefont
		{D.}~\bibnamefont {Bonneau}}, \bibinfo {author} {\bibfnamefont {J.~W.}\
		\bibnamefont {Silverstone}}, \bibinfo {author} {\bibfnamefont
		{Q.}~\bibnamefont {Gong}}, \bibinfo {author} {\bibfnamefont {A.}~\bibnamefont
		{Ac{\'{\i}}n}}, \bibinfo {author} {\bibfnamefont {K.}~\bibnamefont
		{Rottwitt}}, \bibinfo {author} {\bibfnamefont {L.~K.}\ \bibnamefont
		{Oxenl{\o}we}}, \bibinfo {author} {\bibfnamefont {J.~L.}\ \bibnamefont
		{O'Brien}}, \bibinfo {author} {\bibfnamefont {A.}~\bibnamefont {Laing}}, \
		and\ \bibinfo {author} {\bibfnamefont {M.~G.}\ \bibnamefont {Thompson}},\
		}\href {\doibase 10.1126/science.aar7053} {\bibfield  {journal} {\bibinfo
		{journal} {Science (New York, N.Y.)}\ }\textbf {\bibinfo {volume} {360}},\
		\bibinfo {pages} {285} (\bibinfo {year} {2018}{\natexlab{a}})}\BibitemShut
		{NoStop}%
		\bibitem [{\citenamefont {Erhard}\ \emph {et~al.}(2018)\citenamefont {Erhard},
		\citenamefont {Fickler}, \citenamefont {Krenn},\ and\ \citenamefont
		{Zeilinger}}]{Erhard2018}%
		\BibitemOpen
		\bibfield  {author} {\bibinfo {author} {\bibfnamefont {M.}~\bibnamefont
		{Erhard}}, \bibinfo {author} {\bibfnamefont {R.}~\bibnamefont {Fickler}},
		\bibinfo {author} {\bibfnamefont {M.}~\bibnamefont {Krenn}}, \ and\ \bibinfo
		{author} {\bibfnamefont {A.}~\bibnamefont {Zeilinger}},\ }\href {\doibase
		10.1038/lsa.2017.146} {\bibfield  {journal} {\bibinfo  {journal} {Light:
		Science {\&} Applications}\ }\textbf {\bibinfo {volume} {7}},\ \bibinfo
		{pages} {17146} (\bibinfo {year} {2018})}\BibitemShut {NoStop}%
		\bibitem [{\citenamefont {Kaszlikowski}\ \emph {et~al.}(2000)\citenamefont
		{Kaszlikowski}, \citenamefont {Gnaciński}, \citenamefont {Żukowski},
		\citenamefont {Miklaszewski},\ and\ \citenamefont
		{Zeilinger}}]{Kaszlikowski2000}%
		\BibitemOpen
		\bibfield  {author} {\bibinfo {author} {\bibfnamefont {D.}~\bibnamefont
		{Kaszlikowski}}, \bibinfo {author} {\bibfnamefont {P.}~\bibnamefont
		{Gnaciński}}, \bibinfo {author} {\bibfnamefont {M.}~\bibnamefont
		{Żukowski}}, \bibinfo {author} {\bibfnamefont {W.}~\bibnamefont
		{Miklaszewski}}, \ and\ \bibinfo {author} {\bibfnamefont {A.}~\bibnamefont
		{Zeilinger}},\ }\href {\doibase 10.1103/PhysRevLett.85.4418} {\bibfield
		{journal} {\bibinfo  {journal} {Physical Review Letters}\ }\textbf {\bibinfo
		{volume} {85}},\ \bibinfo {pages} {4418} (\bibinfo {year}
		{2000})}\BibitemShut {NoStop}%
		\bibitem [{\citenamefont {Cerf}\ \emph {et~al.}(2002)\citenamefont {Cerf},
		\citenamefont {Bourennane}, \citenamefont {Karlsson},\ and\ \citenamefont
		{Gisin}}]{Cerf2002}%
		\BibitemOpen
		\bibfield  {author} {\bibinfo {author} {\bibfnamefont {N.~J.}\ \bibnamefont
		{Cerf}}, \bibinfo {author} {\bibfnamefont {M.}~\bibnamefont {Bourennane}},
		\bibinfo {author} {\bibfnamefont {A.}~\bibnamefont {Karlsson}}, \ and\
		\bibinfo {author} {\bibfnamefont {N.}~\bibnamefont {Gisin}},\ }\href
		{\doibase 10.1103/PhysRevLett.88.127902} {\bibfield  {journal} {\bibinfo
		{journal} {Physical Review Letters}\ }\textbf {\bibinfo {volume} {88}},\
		\bibinfo {pages} {127902} (\bibinfo {year} {2002})}\BibitemShut {NoStop}%
		\bibitem [{\citenamefont {Grier}(2003)}]{Grier2003}%
		\BibitemOpen
		\bibfield  {author} {\bibinfo {author} {\bibfnamefont {D.~G.}\ \bibnamefont
		{Grier}},\ }\href {\doibase 10.1038/nature01935} {\bibfield  {journal}
		{\bibinfo  {journal} {Nature}\ }\textbf {\bibinfo {volume} {424}},\ \bibinfo
		{pages} {810} (\bibinfo {year} {2003})}\BibitemShut {NoStop}%
		\bibitem [{\citenamefont {Padgett}\ and\ \citenamefont
		{Bowman}(2011)}]{Padgett2011}%
		\BibitemOpen
		\bibfield  {author} {\bibinfo {author} {\bibfnamefont {M.}~\bibnamefont
		{Padgett}}\ and\ \bibinfo {author} {\bibfnamefont {R.}~\bibnamefont
		{Bowman}},\ }\href {\doibase 10.1038/nphoton.2011.81} {\bibfield  {journal}
		{\bibinfo  {journal} {Nature Photonics}\ }\textbf {\bibinfo {volume} {5}},\
		\bibinfo {pages} {343} (\bibinfo {year} {2011})}\BibitemShut {NoStop}%
		\bibitem [{\citenamefont {Gao}\ \emph {et~al.}(2017)\citenamefont {Gao},
		\citenamefont {Ding}, \citenamefont {Nieto-Vesperinas}, \citenamefont {Ding},
		\citenamefont {Rahman}, \citenamefont {Zhang}, \citenamefont {Lim},\ and\
		\citenamefont {Qiu}}]{Gao2017}%
		\BibitemOpen
		\bibfield  {author} {\bibinfo {author} {\bibfnamefont {D.}~\bibnamefont
		{Gao}}, \bibinfo {author} {\bibfnamefont {W.}~\bibnamefont {Ding}}, \bibinfo
		{author} {\bibfnamefont {M.}~\bibnamefont {Nieto-Vesperinas}}, \bibinfo
		{author} {\bibfnamefont {X.}~\bibnamefont {Ding}}, \bibinfo {author}
		{\bibfnamefont {M.}~\bibnamefont {Rahman}}, \bibinfo {author} {\bibfnamefont
		{T.}~\bibnamefont {Zhang}}, \bibinfo {author} {\bibfnamefont
		{C.}~\bibnamefont {Lim}}, \ and\ \bibinfo {author} {\bibfnamefont {C.-W.}\
		\bibnamefont {Qiu}},\ }\href {\doibase 10.1038/lsa.2017.39} {\bibfield
		{journal} {\bibinfo  {journal} {Light: Science {\&} Applications}\ }\textbf
		{\bibinfo {volume} {6}},\ \bibinfo {pages} {e17039} (\bibinfo {year}
		{2017})}\BibitemShut {NoStop}%
		\bibitem [{\citenamefont {Aspelmeyer}\ \emph {et~al.}(2014)\citenamefont
		{Aspelmeyer}, \citenamefont {Kippenberg},\ and\ \citenamefont
		{Marquardt}}]{Aspelmeyer2014}%
		\BibitemOpen
		\bibfield  {author} {\bibinfo {author} {\bibfnamefont {M.}~\bibnamefont
		{Aspelmeyer}}, \bibinfo {author} {\bibfnamefont {T.~J.}\ \bibnamefont
		{Kippenberg}}, \ and\ \bibinfo {author} {\bibfnamefont {F.}~\bibnamefont
		{Marquardt}},\ }\href {\doibase 10.1103/RevModPhys.86.1391} {\bibfield
		{journal} {\bibinfo  {journal} {Reviews of Modern Physics}\ }\textbf
		{\bibinfo {volume} {86}},\ \bibinfo {pages} {1391} (\bibinfo {year}
		{2014})}\BibitemShut {NoStop}%
		\bibitem [{\citenamefont {Collins}\ \emph {et~al.}(2002)\citenamefont
		{Collins}, \citenamefont {Gisin}, \citenamefont {Linden}, \citenamefont
		{Massar},\ and\ \citenamefont {Popescu}}]{Collins2002}%
		\BibitemOpen
		\bibfield  {author} {\bibinfo {author} {\bibfnamefont {D.}~\bibnamefont
		{Collins}}, \bibinfo {author} {\bibfnamefont {N.}~\bibnamefont {Gisin}},
		\bibinfo {author} {\bibfnamefont {N.}~\bibnamefont {Linden}}, \bibinfo
		{author} {\bibfnamefont {S.}~\bibnamefont {Massar}}, \ and\ \bibinfo {author}
		{\bibfnamefont {S.}~\bibnamefont {Popescu}},\ }\href {\doibase
		10.1103/PhysRevLett.88.040404} {\bibfield  {journal} {\bibinfo  {journal}
		{Physical Review Letters}\ }\textbf {\bibinfo {volume} {88}},\ \bibinfo
		{pages} {040404} (\bibinfo {year} {2002})}\BibitemShut {NoStop}%
		\bibitem [{\citenamefont {Fickler}\ \emph {et~al.}(2012)\citenamefont
		{Fickler}, \citenamefont {Lapkiewicz}, \citenamefont {Plick}, \citenamefont
		{Krenn}, \citenamefont {Schaeff}, \citenamefont {Ramelow},\ and\
		\citenamefont {Zeilinger}}]{Fickler2012}%
		\BibitemOpen
		\bibfield  {author} {\bibinfo {author} {\bibfnamefont {R.}~\bibnamefont
		{Fickler}}, \bibinfo {author} {\bibfnamefont {R.}~\bibnamefont {Lapkiewicz}},
		\bibinfo {author} {\bibfnamefont {W.~N.}\ \bibnamefont {Plick}}, \bibinfo
		{author} {\bibfnamefont {M.}~\bibnamefont {Krenn}}, \bibinfo {author}
		{\bibfnamefont {C.}~\bibnamefont {Schaeff}}, \bibinfo {author} {\bibfnamefont
		{S.}~\bibnamefont {Ramelow}}, \ and\ \bibinfo {author} {\bibfnamefont
		{A.}~\bibnamefont {Zeilinger}},\ }\href {\doibase 10.1126/science.1227193}
		{\bibfield  {journal} {\bibinfo  {journal} {Science (New York, N.Y.)}\
		}\textbf {\bibinfo {volume} {338}},\ \bibinfo {pages} {640} (\bibinfo {year}
		{2012})}\BibitemShut {NoStop}%
		\bibitem [{\citenamefont {Simpson}\ \emph {et~al.}(1997)\citenamefont
		{Simpson}, \citenamefont {Dholakia}, \citenamefont {Allen},\ and\
		\citenamefont {Padgett}}]{Simpson1997}%
		\BibitemOpen
		\bibfield  {author} {\bibinfo {author} {\bibfnamefont {N.~B.}\ \bibnamefont
		{Simpson}}, \bibinfo {author} {\bibfnamefont {K.}~\bibnamefont {Dholakia}},
		\bibinfo {author} {\bibfnamefont {L.}~\bibnamefont {Allen}}, \ and\ \bibinfo
		{author} {\bibfnamefont {M.~J.}\ \bibnamefont {Padgett}},\ }\href {\doibase
		10.1364/OL.22.000052} {\bibfield  {journal} {\bibinfo  {journal} {Optics
		Letters}\ }\textbf {\bibinfo {volume} {22}},\ \bibinfo {pages} {52} (\bibinfo
		{year} {1997})}\BibitemShut {NoStop}%
		\bibitem [{\citenamefont {Schemmel}\ \emph {et~al.}(2014)\citenamefont
		{Schemmel}, \citenamefont {Pisano},\ and\ \citenamefont
		{Maffei}}]{Schemmel2014}%
		\BibitemOpen
		\bibfield  {author} {\bibinfo {author} {\bibfnamefont {P.}~\bibnamefont
		{Schemmel}}, \bibinfo {author} {\bibfnamefont {G.}~\bibnamefont {Pisano}}, \
		and\ \bibinfo {author} {\bibfnamefont {B.}~\bibnamefont {Maffei}},\ }\href
		{\doibase 10.1364/OE.22.014712} {\bibfield  {journal} {\bibinfo  {journal}
		{Optics Express}\ }\textbf {\bibinfo {volume} {22}},\ \bibinfo {pages}
		{14712} (\bibinfo {year} {2014})}\BibitemShut {NoStop}%
		\bibitem [{\citenamefont {Mirhosseini}\ \emph {et~al.}(2013)\citenamefont
		{Mirhosseini}, \citenamefont {Maga{\~{n}}a-Loaiza}, \citenamefont {Chen},
		\citenamefont {Rodenburg}, \citenamefont {Malik},\ and\ \citenamefont
		{Boyd}}]{Mirhosseini2013}%
		\BibitemOpen
		\bibfield  {author} {\bibinfo {author} {\bibfnamefont {M.}~\bibnamefont
		{Mirhosseini}}, \bibinfo {author} {\bibfnamefont {O.~S.}\ \bibnamefont
		{Maga{\~{n}}a-Loaiza}}, \bibinfo {author} {\bibfnamefont {C.}~\bibnamefont
		{Chen}}, \bibinfo {author} {\bibfnamefont {B.}~\bibnamefont {Rodenburg}},
		\bibinfo {author} {\bibfnamefont {M.}~\bibnamefont {Malik}}, \ and\ \bibinfo
		{author} {\bibfnamefont {R.~W.}\ \bibnamefont {Boyd}},\ }\href {\doibase
		10.1364/OE.21.030196} {\bibfield  {journal} {\bibinfo  {journal} {Optics
		Express}\ }\textbf {\bibinfo {volume} {21}},\ \bibinfo {pages} {30196}
		(\bibinfo {year} {2013})}\BibitemShut {NoStop}%
		\bibitem [{\citenamefont {Lin}\ \emph {et~al.}(2014)\citenamefont {Lin},
		\citenamefont {Fan}, \citenamefont {Hasman},\ and\ \citenamefont
		{Brongersma}}]{Lin2014}%
		\BibitemOpen
		\bibfield  {author} {\bibinfo {author} {\bibfnamefont {D.}~\bibnamefont
		{Lin}}, \bibinfo {author} {\bibfnamefont {P.}~\bibnamefont {Fan}}, \bibinfo
		{author} {\bibfnamefont {E.}~\bibnamefont {Hasman}}, \ and\ \bibinfo {author}
		{\bibfnamefont {M.~L.}\ \bibnamefont {Brongersma}},\ }\href {\doibase
		10.1126/science.1253213} {\bibfield  {journal} {\bibinfo  {journal} {Science
		(New York, N.Y.)}\ }\textbf {\bibinfo {volume} {345}},\ \bibinfo {pages}
		{298} (\bibinfo {year} {2014})}\BibitemShut {NoStop}%
		\bibitem [{\citenamefont {Arbabi}\ \emph {et~al.}(2015)\citenamefont {Arbabi},
		\citenamefont {Horie}, \citenamefont {Bagheri},\ and\ \citenamefont
		{Faraon}}]{Arbabi2015}%
		\BibitemOpen
		\bibfield  {author} {\bibinfo {author} {\bibfnamefont {A.}~\bibnamefont
		{Arbabi}}, \bibinfo {author} {\bibfnamefont {Y.}~\bibnamefont {Horie}},
		\bibinfo {author} {\bibfnamefont {M.}~\bibnamefont {Bagheri}}, \ and\
		\bibinfo {author} {\bibfnamefont {A.}~\bibnamefont {Faraon}},\ }\href
		{\doibase 10.1038/nnano.2015.186} {\bibfield  {journal} {\bibinfo  {journal}
		{Nature Nanotechnology}\ }\textbf {\bibinfo {volume} {10}},\ \bibinfo {pages}
		{937} (\bibinfo {year} {2015})}\BibitemShut {NoStop}%
		\bibitem [{\citenamefont {Devlin}\ \emph {et~al.}(2017)\citenamefont {Devlin},
		\citenamefont {Ambrosio}, \citenamefont {Rubin}, \citenamefont {Mueller},\
		and\ \citenamefont {Capasso}}]{Devlin2017}%
		\BibitemOpen
		\bibfield  {author} {\bibinfo {author} {\bibfnamefont {R.~C.}\ \bibnamefont
		{Devlin}}, \bibinfo {author} {\bibfnamefont {A.}~\bibnamefont {Ambrosio}},
		\bibinfo {author} {\bibfnamefont {N.~A.}\ \bibnamefont {Rubin}}, \bibinfo
		{author} {\bibfnamefont {J.~P.~B.}\ \bibnamefont {Mueller}}, \ and\ \bibinfo
		{author} {\bibfnamefont {F.}~\bibnamefont {Capasso}},\ }\href {\doibase
		10.1126/science.aao5392} {\bibfield  {journal} {\bibinfo  {journal} {Science
		(New York, N.Y.)}\ }\textbf {\bibinfo {volume} {358}},\ \bibinfo {pages}
		{896} (\bibinfo {year} {2017})}\BibitemShut {NoStop}%
		\bibitem [{\citenamefont {Xu}\ \emph {et~al.}(2018)\citenamefont {Xu},
		\citenamefont {Sun}, \citenamefont {Frantz}, \citenamefont {Shalaev},
		\citenamefont {Walasik}, \citenamefont {Pandey}, \citenamefont {Myers},
		\citenamefont {Bekele}, \citenamefont {Tsukernik}, \citenamefont {Sanghera},\
		and\ \citenamefont {Litchinitser}}]{Xu2018}%
		\BibitemOpen
		\bibfield  {author} {\bibinfo {author} {\bibfnamefont {Y.}~\bibnamefont
		{Xu}}, \bibinfo {author} {\bibfnamefont {J.}~\bibnamefont {Sun}}, \bibinfo
		{author} {\bibfnamefont {J.}~\bibnamefont {Frantz}}, \bibinfo {author}
		{\bibfnamefont {M.~I.}\ \bibnamefont {Shalaev}}, \bibinfo {author}
		{\bibfnamefont {W.}~\bibnamefont {Walasik}}, \bibinfo {author} {\bibfnamefont
		{A.}~\bibnamefont {Pandey}}, \bibinfo {author} {\bibfnamefont {J.~D.}\
		\bibnamefont {Myers}}, \bibinfo {author} {\bibfnamefont {R.~Y.}\ \bibnamefont
		{Bekele}}, \bibinfo {author} {\bibfnamefont {A.}~\bibnamefont {Tsukernik}},
		\bibinfo {author} {\bibfnamefont {J.~S.}\ \bibnamefont {Sanghera}}, \ and\
		\bibinfo {author} {\bibfnamefont {N.~M.}\ \bibnamefont {Litchinitser}},\
		}\href {http://arxiv.org/abs/1805.07327} {\  (\bibinfo {year} {2018})},\
		\Eprint {http://arxiv.org/abs/1805.07327} {arXiv:1805.07327} \BibitemShut
		{NoStop}%
		\bibitem [{\citenamefont {Marrucci}\ \emph {et~al.}(2006)\citenamefont
		{Marrucci}, \citenamefont {Manzo},\ and\ \citenamefont
		{Paparo}}]{Marrucci2006}%
		\BibitemOpen
		\bibfield  {author} {\bibinfo {author} {\bibfnamefont {L.}~\bibnamefont
		{Marrucci}}, \bibinfo {author} {\bibfnamefont {C.}~\bibnamefont {Manzo}}, \
		and\ \bibinfo {author} {\bibfnamefont {D.}~\bibnamefont {Paparo}},\ }\href
		{\doibase 10.1103/PhysRevLett.96.163905} {\bibfield  {journal} {\bibinfo
		{journal} {Physical Review Letters}\ }\textbf {\bibinfo {volume} {96}},\
		\bibinfo {pages} {163905} (\bibinfo {year} {2006})}\BibitemShut {NoStop}%
		\bibitem [{\citenamefont {Karimi}\ \emph {et~al.}(2009)\citenamefont {Karimi},
		\citenamefont {Piccirillo}, \citenamefont {Nagali}, \citenamefont
		{Marrucci},\ and\ \citenamefont {Santamato}}]{Karimi2009}%
		\BibitemOpen
		\bibfield  {author} {\bibinfo {author} {\bibfnamefont {E.}~\bibnamefont
		{Karimi}}, \bibinfo {author} {\bibfnamefont {B.}~\bibnamefont {Piccirillo}},
		\bibinfo {author} {\bibfnamefont {E.}~\bibnamefont {Nagali}}, \bibinfo
		{author} {\bibfnamefont {L.}~\bibnamefont {Marrucci}}, \ and\ \bibinfo
		{author} {\bibfnamefont {E.}~\bibnamefont {Santamato}},\ }\href {\doibase
		10.1063/1.3154549} {\bibfield  {journal} {\bibinfo  {journal} {Applied
		Physics Letters}\ }\textbf {\bibinfo {volume} {94}},\ \bibinfo {pages}
		{231124} (\bibinfo {year} {2009})}\BibitemShut {NoStop}%
		\bibitem [{\citenamefont {Dall}\ \emph {et~al.}(2014)\citenamefont {Dall},
		\citenamefont {Fraser}, \citenamefont {Desyatnikov}, \citenamefont {Li},
		\citenamefont {Brodbeck}, \citenamefont {Kamp}, \citenamefont {Schneider},
		\citenamefont {H{\"{o}}fling},\ and\ \citenamefont {Ostrovskaya}}]{Dall2014}%
		\BibitemOpen
		\bibfield  {author} {\bibinfo {author} {\bibfnamefont {R.}~\bibnamefont
		{Dall}}, \bibinfo {author} {\bibfnamefont {M.~D.}\ \bibnamefont {Fraser}},
		\bibinfo {author} {\bibfnamefont {A.~S.}\ \bibnamefont {Desyatnikov}},
		\bibinfo {author} {\bibfnamefont {G.}~\bibnamefont {Li}}, \bibinfo {author}
		{\bibfnamefont {S.}~\bibnamefont {Brodbeck}}, \bibinfo {author}
		{\bibfnamefont {M.}~\bibnamefont {Kamp}}, \bibinfo {author} {\bibfnamefont
		{C.}~\bibnamefont {Schneider}}, \bibinfo {author} {\bibfnamefont
		{S.}~\bibnamefont {H{\"{o}}fling}}, \ and\ \bibinfo {author} {\bibfnamefont
		{E.~A.}\ \bibnamefont {Ostrovskaya}},\ }\href {\doibase
		10.1103/PhysRevLett.113.200404} {\bibfield  {journal} {\bibinfo  {journal}
		{Physical Review Letters}\ }\textbf {\bibinfo {volume} {113}},\ \bibinfo
		{pages} {200404} (\bibinfo {year} {2014})}\BibitemShut {NoStop}%
		\bibitem [{\citenamefont {Gao}\ \emph {et~al.}(2018)\citenamefont {Gao},
		\citenamefont {Li}, \citenamefont {Estrecho}, \citenamefont {Liew},
		\citenamefont {Comber-Todd}, \citenamefont {Nalitov}, \citenamefont {Steger},
		\citenamefont {West}, \citenamefont {Pfeiffer}, \citenamefont {Snoke},
		\citenamefont {Kavokin}, \citenamefont {Truscott},\ and\ \citenamefont
		{Ostrovskaya}}]{Gao2018}%
		\BibitemOpen
		\bibfield  {author} {\bibinfo {author} {\bibfnamefont {T.}~\bibnamefont
		{Gao}}, \bibinfo {author} {\bibfnamefont {G.}~\bibnamefont {Li}}, \bibinfo
		{author} {\bibfnamefont {E.}~\bibnamefont {Estrecho}}, \bibinfo {author}
		{\bibfnamefont {T.}~\bibnamefont {Liew}}, \bibinfo {author} {\bibfnamefont
		{D.}~\bibnamefont {Comber-Todd}}, \bibinfo {author} {\bibfnamefont
		{A.}~\bibnamefont {Nalitov}}, \bibinfo {author} {\bibfnamefont
		{M.}~\bibnamefont {Steger}}, \bibinfo {author} {\bibfnamefont
		{K.}~\bibnamefont {West}}, \bibinfo {author} {\bibfnamefont {L.}~\bibnamefont
		{Pfeiffer}}, \bibinfo {author} {\bibfnamefont {D.}~\bibnamefont {Snoke}},
		\bibinfo {author} {\bibfnamefont {A.}~\bibnamefont {Kavokin}}, \bibinfo
		{author} {\bibfnamefont {A.}~\bibnamefont {Truscott}}, \ and\ \bibinfo
		{author} {\bibfnamefont {E.}~\bibnamefont {Ostrovskaya}},\ }\href {\doibase
		10.1103/PhysRevLett.120.065301} {\bibfield  {journal} {\bibinfo  {journal}
		{Physical Review Letters}\ }\textbf {\bibinfo {volume} {120}},\ \bibinfo
		{pages} {065301} (\bibinfo {year} {2018})}\BibitemShut {NoStop}%
		\bibitem [{\citenamefont {Miao}\ \emph {et~al.}(2016)\citenamefont {Miao},
		\citenamefont {Zhang}, \citenamefont {Sun}, \citenamefont {Walasik},
		\citenamefont {Longhi}, \citenamefont {Litchinitser},\ and\ \citenamefont
		{Feng}}]{Miao2016}%
		\BibitemOpen
		\bibfield  {author} {\bibinfo {author} {\bibfnamefont {P.}~\bibnamefont
		{Miao}}, \bibinfo {author} {\bibfnamefont {Z.}~\bibnamefont {Zhang}},
		\bibinfo {author} {\bibfnamefont {J.}~\bibnamefont {Sun}}, \bibinfo {author}
		{\bibfnamefont {W.}~\bibnamefont {Walasik}}, \bibinfo {author} {\bibfnamefont
		{S.}~\bibnamefont {Longhi}}, \bibinfo {author} {\bibfnamefont {N.~M.}\
		\bibnamefont {Litchinitser}}, \ and\ \bibinfo {author} {\bibfnamefont
		{L.}~\bibnamefont {Feng}},\ }\href {\doibase 10.1126/science.aaf8533}
		{\bibfield  {journal} {\bibinfo  {journal} {Science (New York, N.Y.)}\
		}\textbf {\bibinfo {volume} {353}},\ \bibinfo {pages} {464} (\bibinfo {year}
		{2016})}\BibitemShut {NoStop}%
		\bibitem [{\citenamefont {Peng}\ \emph {et~al.}(2016)\citenamefont {Peng},
		\citenamefont {{\"{O}}zdemir}, \citenamefont {Liertzer}, \citenamefont
		{Chen}, \citenamefont {Kramer}, \citenamefont {Yılmaz}, \citenamefont
		{Wiersig}, \citenamefont {Rotter},\ and\ \citenamefont {Yang}}]{Peng2016}%
		\BibitemOpen
		\bibfield  {author} {\bibinfo {author} {\bibfnamefont {B.}~\bibnamefont
		{Peng}}, \bibinfo {author} {\bibfnamefont {a.~K.}\ \bibnamefont
		{{\"{O}}zdemir}}, \bibinfo {author} {\bibfnamefont {M.}~\bibnamefont
		{Liertzer}}, \bibinfo {author} {\bibfnamefont {W.}~\bibnamefont {Chen}},
		\bibinfo {author} {\bibfnamefont {J.}~\bibnamefont {Kramer}}, \bibinfo
		{author} {\bibfnamefont {H.}~\bibnamefont {Yılmaz}}, \bibinfo {author}
		{\bibfnamefont {J.}~\bibnamefont {Wiersig}}, \bibinfo {author} {\bibfnamefont
		{S.}~\bibnamefont {Rotter}}, \ and\ \bibinfo {author} {\bibfnamefont
		{L.}~\bibnamefont {Yang}},\ }\href {\doibase 10.1073/pnas.1603318113}
		{\bibfield  {journal} {\bibinfo  {journal} {Proceedings of the National
		Academy of Sciences of the United States of America}\ }\textbf {\bibinfo
		{volume} {113}},\ \bibinfo {pages} {6845} (\bibinfo {year}
		{2016})}\BibitemShut {NoStop}%
		\bibitem [{\citenamefont {Sala}\ \emph {et~al.}(2015)\citenamefont {Sala},
		\citenamefont {Solnyshkov}, \citenamefont {Carusotto}, \citenamefont
		{Jacqmin}, \citenamefont {Lema{\^{\i}}tre}, \citenamefont {Ter{\c{c}}as},
		\citenamefont {Nalitov}, \citenamefont {Abbarchi}, \citenamefont {Galopin},
		\citenamefont {Sagnes}, \citenamefont {Bloch}, \citenamefont {Malpuech},\
		and\ \citenamefont {Amo}}]{Sala2015}%
		\BibitemOpen
		\bibfield  {author} {\bibinfo {author} {\bibfnamefont {V.}~\bibnamefont
		{Sala}}, \bibinfo {author} {\bibfnamefont {D.}~\bibnamefont {Solnyshkov}},
		\bibinfo {author} {\bibfnamefont {I.}~\bibnamefont {Carusotto}}, \bibinfo
		{author} {\bibfnamefont {T.}~\bibnamefont {Jacqmin}}, \bibinfo {author}
		{\bibfnamefont {A.}~\bibnamefont {Lema{\^{\i}}tre}}, \bibinfo {author}
		{\bibfnamefont {H.}~\bibnamefont {Ter{\c{c}}as}}, \bibinfo {author}
		{\bibfnamefont {A.}~\bibnamefont {Nalitov}}, \bibinfo {author} {\bibfnamefont
		{M.}~\bibnamefont {Abbarchi}}, \bibinfo {author} {\bibfnamefont
		{E.}~\bibnamefont {Galopin}}, \bibinfo {author} {\bibfnamefont
		{I.}~\bibnamefont {Sagnes}}, \bibinfo {author} {\bibfnamefont
		{J.}~\bibnamefont {Bloch}}, \bibinfo {author} {\bibfnamefont
		{G.}~\bibnamefont {Malpuech}}, \ and\ \bibinfo {author} {\bibfnamefont
		{A.}~\bibnamefont {Amo}},\ }\href {\doibase 10.1103/PhysRevX.5.011034}
		{\bibfield  {journal} {\bibinfo  {journal} {Physical Review X}\ }\textbf
		{\bibinfo {volume} {5}},\ \bibinfo {pages} {011034} (\bibinfo {year}
		{2015})}\BibitemShut {NoStop}%
		\bibitem [{\citenamefont {Cardano}\ and\ \citenamefont
		{Marrucci}(2015)}]{Cardano2015}%
		\BibitemOpen
		\bibfield  {author} {\bibinfo {author} {\bibfnamefont {F.}~\bibnamefont
		{Cardano}}\ and\ \bibinfo {author} {\bibfnamefont {L.}~\bibnamefont
		{Marrucci}},\ }\href {\doibase 10.1038/nphoton.2015.232} {\bibfield
		{journal} {\bibinfo  {journal} {Nature Photonics}\ }\textbf {\bibinfo
		{volume} {9}},\ \bibinfo {pages} {776} (\bibinfo {year} {2015})}\BibitemShut
		{NoStop}%
		\bibitem [{\citenamefont {{Michaelis de Vasconcellos}}\ \emph
		{et~al.}(2011)\citenamefont {{Michaelis de Vasconcellos}}, \citenamefont
		{Calvar}, \citenamefont {Dousse}, \citenamefont {Suffczyński}, \citenamefont
		{Dupuis}, \citenamefont {Lema{\^{\i}}tre}, \citenamefont {Sagnes},
		\citenamefont {Bloch}, \citenamefont {Voisin},\ and\ \citenamefont
		{Senellart}}]{MichaelisdeVasconcellos2011}%
		\BibitemOpen
		\bibfield  {author} {\bibinfo {author} {\bibfnamefont {S.}~\bibnamefont
		{{Michaelis de Vasconcellos}}}, \bibinfo {author} {\bibfnamefont
		{A.}~\bibnamefont {Calvar}}, \bibinfo {author} {\bibfnamefont
		{A.}~\bibnamefont {Dousse}}, \bibinfo {author} {\bibfnamefont
		{J.}~\bibnamefont {Suffczyński}}, \bibinfo {author} {\bibfnamefont
		{N.}~\bibnamefont {Dupuis}}, \bibinfo {author} {\bibfnamefont
		{A.}~\bibnamefont {Lema{\^{\i}}tre}}, \bibinfo {author} {\bibfnamefont
		{I.}~\bibnamefont {Sagnes}}, \bibinfo {author} {\bibfnamefont
		{J.}~\bibnamefont {Bloch}}, \bibinfo {author} {\bibfnamefont
		{P.}~\bibnamefont {Voisin}}, \ and\ \bibinfo {author} {\bibfnamefont
		{P.}~\bibnamefont {Senellart}},\ }\href {\doibase 10.1063/1.3632111}
		{\bibfield  {journal} {\bibinfo  {journal} {Applied Physics Letters}\
		}\textbf {\bibinfo {volume} {99}},\ \bibinfo {pages} {101103} (\bibinfo
		{year} {2011})}\BibitemShut {NoStop}%
		\bibitem [{\citenamefont {Wang}\ \emph
		{et~al.}(2018{\natexlab{b}})\citenamefont {Wang}, \citenamefont {Chernikov},
		\citenamefont {Glazov}, \citenamefont {Heinz}, \citenamefont {Marie},
		\citenamefont {Amand},\ and\ \citenamefont {Urbaszek}}]{Wang2018a}%
		\BibitemOpen
		\bibfield  {author} {\bibinfo {author} {\bibfnamefont {G.}~\bibnamefont
		{Wang}}, \bibinfo {author} {\bibfnamefont {A.}~\bibnamefont {Chernikov}},
		\bibinfo {author} {\bibfnamefont {M.~M.}\ \bibnamefont {Glazov}}, \bibinfo
		{author} {\bibfnamefont {T.~F.}\ \bibnamefont {Heinz}}, \bibinfo {author}
		{\bibfnamefont {X.}~\bibnamefont {Marie}}, \bibinfo {author} {\bibfnamefont
		{T.}~\bibnamefont {Amand}}, \ and\ \bibinfo {author} {\bibfnamefont
		{B.}~\bibnamefont {Urbaszek}},\ }\href {\doibase
		10.1103/RevModPhys.90.021001} {\bibfield  {journal} {\bibinfo  {journal}
		{Reviews of Modern Physics}\ }\textbf {\bibinfo {volume} {90}},\ \bibinfo
		{pages} {021001} (\bibinfo {year} {2018}{\natexlab{b}})}\BibitemShut
		{NoStop}%
		\bibitem [{\citenamefont {Allain}\ \emph {et~al.}(2015)\citenamefont {Allain},
		\citenamefont {Kang}, \citenamefont {Banerjee},\ and\ \citenamefont
		{Kis}}]{Allain2015}%
		\BibitemOpen
		\bibfield  {author} {\bibinfo {author} {\bibfnamefont {A.}~\bibnamefont
		{Allain}}, \bibinfo {author} {\bibfnamefont {J.}~\bibnamefont {Kang}},
		\bibinfo {author} {\bibfnamefont {K.}~\bibnamefont {Banerjee}}, \ and\
		\bibinfo {author} {\bibfnamefont {A.}~\bibnamefont {Kis}},\ }\href {\doibase
		10.1038/nmat4452} {\bibfield  {journal} {\bibinfo  {journal} {Nature
		Materials}\ }\textbf {\bibinfo {volume} {14}},\ \bibinfo {pages} {1195}
		(\bibinfo {year} {2015})}\BibitemShut {NoStop}%
		\bibitem [{\citenamefont {Wolf}\ \emph {et~al.}(2001)\citenamefont {Wolf},
		\citenamefont {Awschalom}, \citenamefont {Buhrman}, \citenamefont {Daughton},
		\citenamefont {von Moln{\'{a}}r}, \citenamefont {Roukes}, \citenamefont
		{Chtchelkanova},\ and\ \citenamefont {Treger}}]{Wolf2001}%
		\BibitemOpen
		\bibfield  {author} {\bibinfo {author} {\bibfnamefont {S.~A.}\ \bibnamefont
		{Wolf}}, \bibinfo {author} {\bibfnamefont {D.~D.}\ \bibnamefont {Awschalom}},
		\bibinfo {author} {\bibfnamefont {R.~A.}\ \bibnamefont {Buhrman}}, \bibinfo
		{author} {\bibfnamefont {J.~M.}\ \bibnamefont {Daughton}}, \bibinfo {author}
		{\bibfnamefont {S.}~\bibnamefont {von Moln{\'{a}}r}}, \bibinfo {author}
		{\bibfnamefont {M.~L.}\ \bibnamefont {Roukes}}, \bibinfo {author}
		{\bibfnamefont {A.~Y.}\ \bibnamefont {Chtchelkanova}}, \ and\ \bibinfo
		{author} {\bibfnamefont {D.~M.}\ \bibnamefont {Treger}},\ }\href {\doibase
		10.1126/science.1065389} {\bibfield  {journal} {\bibinfo  {journal} {Science
		(New York, N.Y.)}\ }\textbf {\bibinfo {volume} {294}},\ \bibinfo {pages}
		{1488} (\bibinfo {year} {2001})}\BibitemShut {NoStop}%
		\bibitem [{\citenamefont {Fenton}\ \emph {et~al.}(2018)\citenamefont {Fenton},
		\citenamefont {Khan}, \citenamefont {Solano}, \citenamefont {Orozco},\ and\
		\citenamefont {Fatemi}}]{Fenton2018}%
		\BibitemOpen
		\bibfield  {author} {\bibinfo {author} {\bibfnamefont {E.~F.}\ \bibnamefont
		{Fenton}}, \bibinfo {author} {\bibfnamefont {A.}~\bibnamefont {Khan}},
		\bibinfo {author} {\bibfnamefont {P.}~\bibnamefont {Solano}}, \bibinfo
		{author} {\bibfnamefont {L.~A.}\ \bibnamefont {Orozco}}, \ and\ \bibinfo
		{author} {\bibfnamefont {F.~K.}\ \bibnamefont {Fatemi}},\ }\href {\doibase
		10.1364/OL.43.001534} {\bibfield  {journal} {\bibinfo  {journal} {Optics
		Letters}\ }\textbf {\bibinfo {volume} {43}},\ \bibinfo {pages} {1534}
		(\bibinfo {year} {2018})}\BibitemShut {NoStop}%
		\bibitem [{\citenamefont {Fong}\ \emph {et~al.}(2018)\citenamefont {Fong},
		\citenamefont {Ota}, \citenamefont {Iwamoto},\ and\ \citenamefont
		{Arakawa}}]{Fong2018}%
		\BibitemOpen
		\bibfield  {author} {\bibinfo {author} {\bibfnamefont {C.~F.}\ \bibnamefont
		{Fong}}, \bibinfo {author} {\bibfnamefont {Y.}~\bibnamefont {Ota}}, \bibinfo
		{author} {\bibfnamefont {S.}~\bibnamefont {Iwamoto}}, \ and\ \bibinfo
		{author} {\bibfnamefont {Y.}~\bibnamefont {Arakawa}},\ }\href
		{http://arxiv.org/abs/1806.00223} {\  (\bibinfo {year} {2018})},\ \Eprint
		{http://arxiv.org/abs/1806.00223} {arXiv:1806.00223} \BibitemShut {NoStop}%
		\bibitem [{\citenamefont {Dresselhaus}\ \emph {et~al.}(2008)\citenamefont
		{Dresselhaus}, \citenamefont {Dresselhaus},\ and\ \citenamefont
		{Jorio}}]{Dresselhaus2008}%
		\BibitemOpen
		\bibfield  {author} {\bibinfo {author} {\bibfnamefont {M.~S.}\ \bibnamefont
		{Dresselhaus}}, \bibinfo {author} {\bibfnamefont {G.}~\bibnamefont
		{Dresselhaus}}, \ and\ \bibinfo {author} {\bibfnamefont {A.~A.}\ \bibnamefont
		{Jorio}},\ }\href@noop {} {\emph {\bibinfo {title} {{Group theory :
		application to the physics of condensed matter}}}}\ (\bibinfo  {publisher}
		{Springer-Verlag},\ \bibinfo {year} {2008})\ p.\ \bibinfo {pages}
		{582}\BibitemShut {NoStop}%
		\end{thebibliography}
\end{document}